\documentclass[12pt]{amsart}
\usepackage{setspace}
\usepackage[foot]{amsaddr}
\usepackage{rotating, lscape}
\usepackage{bm}
\usepackage{threeparttable}
\usepackage{longtable}
\usepackage{float}
\usepackage{adjustbox}
\usepackage{amssymb,amsmath,amsfonts,amsthm}
\usepackage{booktabs,colortbl,multirow}
\usepackage{upgreek}
\usepackage{enumitem}
\usepackage{longtable} % for 'longtable' environment
\usepackage{pdflscape} % for 'landscape' environment
\usepackage{algorithm}
\usepackage{algpseudocode}
\usepackage{bbm}
\usepackage{arydshln}
\usepackage{caption}
\usepackage{subcaption}
\usepackage{graphicx}
\usepackage{ragged2e}
\usepackage{setspace}

\usepackage{array}

\newcolumntype{B}{>{\bfseries}c}

\usepackage[firstinits=true,style = authoryear,maxcitenames=5,maxbibnames=99]{biblatex}

\allowdisplaybreaks
\def\0{{\bm{0}}}

\def\be{\begin{eqnarray}}
\def\ee{\end{eqnarray}}

\thanks{Julia Hatamyar, Corresponding Author e-mail: julia.hatamyar@york.ac.uk. All code used in this paper is available at https://github.com/jhatamyar/MLDID}
\IfFileExists{upquote.sty}{\usepackage{upquote}}{}

\begin{document}
	
% \title{Some \textit{Skewed} Results on the Stochastic Frontier Model}
\title[Machine Learning \& Staggered DID]{Machine Learning for Staggered Difference-in-Differences and Dynamic Treatment Effect Heterogeneity}
\date{\today}

\author{Julia Hatamyar}
\author{Noemi Kreif}
\author{Rudi Rocha}
\author{Martin Huber}

\thanks{This work was funded by the UK Medical Research Council (Grant \#: MR/T04487X/1)}

\begin{abstract}
We combine two recently proposed nonparametric difference-in-differences methods, extending them to enable the examination of treatment effect heterogeneity in the staggered adoption setting using machine learning. The proposed method, machine learning difference-in-differences (MLDID), allows for estimation of time-varying conditional average treatment effects on the treated, which can be used to conduct detailed inference on drivers of treatment effect heterogeneity. We perform simulations to evaluate the performance of MLDID and find that it accurately identifies the true predictors of treatment effect heterogeneity. We then use MLDID to evaluate the heterogeneous impacts of Brazil's Family Health Program on infant mortality, and find those in poverty and urban locations experienced the impact of the policy more quickly than other subgroups. 
\end{abstract}

\maketitle

%\section{Introduction}

\section{Introduction}

Difference-in-differences (DID) with staggered treatment adoption, a popular econometric approach to estimating dynamic treatment effects, is a powerful tool for estimating the causal effects of policies or programs in real-world settings. Staggered implementation of treatment can provide a useful quasi-experimental design to estimate the causal effect of a policy by allowing the researcher to control for confounders that change over time. Estimators that rely on the staggered DID design\footnote{See \textcite{baker2022much} for an overview of various estimators for use with staggered DID} require the critical assumption that the joint distribution of treatment and covariates and the timing of treatment implementation are independent \parencite{callaway2021difference}. In other words, the timing of the treatment adoption should not be related to other factors that are also affecting the outcome being examined. This assumption may not hold in many important observational settings, for example when disadvantaged areas are targeted for implementation first.

In this paper, we create a novel and useful extension of existing methods for estimating dynamic treatment effects by combining the nonparametric machine learning-based (ML) DID estimator proposed by  \textcite{lu2019robust}, with a staggered adoption framework as in \textcite{callaway2021difference}. We refer to our extension as  MLDID. One key strength of our method is inherited from \textcite{lu2019robust}, and allows for treatment and timing of treatment to be correlated.\footnote{Using a decomposition inspired by \textcite{robinson1988root}.} This is done by combining ML estimates of various nuisance models (e.g. the outcome model, the propensity score for entering a treatment group, among others unique to the DID setting) in an appropriate way. Our proposed method has another major advantage:  due to the incorporation of ML, it  allows for the investigation of treatment effect heterogeneity in a way that allows for more variables and is more flexible than the traditional approach of adding interaction terms in the analysis or subsetting the data by particular covariate values. By combining ML estimates of nuisance models, and then aggregating them appropriately, we can obtain predictions of dynamic Conditional Average Treatment Effects on the Treated (CATTs) for each unit of observation, (i.e., CATTs for each time period post-treatment). We then propose for these predictions to be used for inference on dynamic treatment effect heterogeneity in a data-driven manner, for example regressed on covariates of interest, inspired by the generic ML framework proposed by \textcite{chernozhukov2018generic}. 

Simulations indicate that the MLDID approach estimates the dynamic average treatment effects with small error. We show that our approach can correctly identify which variables predict treatment effects, and can also obtain good estimates on how the effects of these covariates on the CATT \textit{change over time}. We are the first to empirically demonstrate the application of ML to the staggered DID setting. \textcite{gavrilova2023dynamic} extend the Causal Forest method to a single-treatment timing DID framework and describe its potential use with staggered treatment. Other examples for the use of ML with a single-treatment timing DID include \textcite{chang2020double} and \textcite{zimmert2020efficient} (see Section 2). Yet, we are unaware of other methods to estimate dynamic CATTs for \textit{staggered} DID, allowing researchers to disentangle the effects of covariates on treatment effect heterogeneity in a dynamic way. 

To demonstrate the usefulness of our proposed technique, we apply the method to a case study of the effect of the Brazilian Family Health Programme (FHP) on infant mortality. Our estimates for the average treatment effect on the treated (ATT) are in line with those obtained from previously developed nonparametric methods \parencite{callaway2021difference}, and show a significant reduction in infant mortality in the years following implementation of the FHP. In addition, we use our CATT estimates to study treatment effect heterogeneity and find that drivers of treatment effect heterogeneity change over time. The decrease in infant mortality in later years is likely related to treatment effect heterogeneity in inequality, poverty, and ethnic diversity. This finding suggests that the intended impact of FHP on disadvantaged areas may have taken longer to materialize, however it eventually contributed to a more significant overall reduction in infant mortality among the treated areas over time.

\section{Background}

\subsection{Staggered Treatment Timing Approaches for Panel Data Event Studies}

A staggered DiD design aims to estimate the impacts of policies that have been introduced over time in a staggered way \parencite{bertrand2004much, wooldridge2010econometric}, potentially leaving no units unaffected by the policy by the end of the time period considered. The approach can address confounding  due to  differences in observed and unobserved characteristics of units that receive treatment at different times.  One popular estimation approach accompanying this design has been the traditional two-way fixed effect regression that accounts for staggered entry to treatment by using cross-sectional and time fixed effects, and a binary treatment indicator with leads and lags to capture the estimated effect in each period (this approach is often referred to as an event study). However, several recent studies have outlined the shortcomings of this method \parencite{baker2021much,de2020two,goodman2018difference,sun2020estimating}, specifically  that incorrectly weighting the mean estimates of groups (defined by their treatment introduction time)  can lead to estimates that do not capture the original causal estimand of interest.  Goodman-Bacon (2018) shows that the TWFE estimator is a weighted average of 2x2 DiD estimates where the weights are based on group size and variance in treatment. Groups treated later in the sample period act as controls before treatment begins. When treatment effects change over time, comparisons between these early and late treated groups can lead to negative weighting in the average treatment effect. 

Callaway and Sant'Anna (2020) propose one of the recently developed methods to overcome weighting issues. They calculate a group ATT by time relative to treatment start for each cohort using a nonparametric approach. In contrast to a parametric outcome regression-based approach, this allows for doubly-robust estimators to be used, such as the DRDiD developed by Sant'Anna and Zhao (2020). Such a double-robust approach allows for unbiased estimation of the group ATT under the assumption that at least one of the two underlying models (outcome regression or treatment propensity score) is correctly specified. It is a natural extension to apply ML methods used in other doubly-robust estimation techniques to this setting, as the incorporation of ML methods can reduce bias due to model misspecification.  

There are other proposed methods to correct for the bias in TWFE estimation. de Chaisemartin and D'Haultfoeuille (2020) mainly focus on instantaneous effects, which are not as useful in a health policy evaluation setting (where the policy impact likely evolves over time). Sun and Abraham (2020) propose regression-based estimates of the group average treatment effects in a similar vein to Callaway and Sant'Anna (2020). However, their method requires an unconditional version of the parallel trends assumption while Callaway and Sant'Anna's allows for the assumption to hold after conditioning on covariates in a more specific manner. In a policy evaluation setting such as the one studied in this paper, where  relatively poorer areas are prioritized  for initial policy implementation, a conditional parallel trends assumption is more likely to hold. The advantage of implementing the interaction-weighted group treatment effect estimation of Sun and Abraham (2020) is that their parametric approach allows more flexibly for time-varying covariates to be included in the regression model. The use of (time-varying) post-treatment covariates is ruled out by Callaway and Sant'Anna (2020). As the authors argue, these can be impacted by the treatment \parencite{wooldridge2005violating}. In applied policy evaluation, it may be important to relax this assumption. There may be some covariates which do not change as a result of the treatment over a reasonable time period (e.g.  the age distribution in a municipality), but it may still be important to control for them, as they can correlated with the treatment start.

\subsection{Non-parametric DID}
In parallel work, Zimmert (2018, 2020) and Chang (2020) each develop a double-robust method to estimate the ATT in panel DID settings.\footnote{See \parencite{abadie2005semiparametric} for a discussion of use of semiparametric estimators in the DID setting where the strong stationarity assumption holds. Also see \textcite{d2022nonparametric} who propose a nonparametric difference-in-differences estimator, but for continuous treatments.} Zimmert proposes a double-robust DiD estimator \parencite{zimmert2018difference} and derives efficient influence functions \parencite{zimmert2020efficient} that allow for flexible first-stage estimators, and applies ML to this using a cross-validation procedure. The DR scores proposed by  Zimmert (2020) maintain double robustness properties without depending on complex requirements that use Gateaux differentiation principles (as in Chernozhukov et al. 2018), focussing on a range of score functions that are rate verifiable.  The paper also implies that including extra covariates in addition to those required for satisfying identification assumptions (and also allowing for time-varying ones)  may potentially increase the precision of some estimators. The double-robust, orthogonal method for DID proposed by \textcite{chang2020double} more directly applies the results of Chernozhukov et al. (2018). It also requires the strong stationarity assumption (i.e., that the joint distribution of treatment and covariates remains constant over time) and does not attain the semiparametric efficiency bound. Neither Chang nor Zimmert apply their method to a panel setting (although both derive the estimator for panel data), instead examining cross-sectional applications in their empirical demonstrations. 

Recently, \textcite{gavrilova2023dynamic} extended the Causal Forest method of \textcite{wager2018estimation} to the DID setting for a single treatment time. Their approach applies base-year differencing as well as orthogonalization to the outcome variable, then estimates a Causal Forest on this transformed outcome variable for each time period, resulting in estimates of average and heterogeneous treatment effects (i.e., group ATTs and CATTs, for a single treatment group). The method only relies on satisfying the parallel trends assumption (and not on assuming that treatment assignment is independent of potential outcomes, i.e., the conditional independence assumption). The extension to the staggered treatment timing setting is described but not implemented. 

 Finally, Lu et al (2019) propose a nonparametric DiD approach that requires no additional assumptions beyond conditional parallel trends \parencite{lu2019robust}, i.e., relaxing the strong stationarity assumption.\footnote{\parencite{hong2013measuring} addresses a scenario where the composition of the treatment group changes over time using a matching technique.} They also propose a heterogeneous treatment effect estimator for a certain type of conditional treatment effect (the CATT). The authors only deal with repeated cross-sections in their empirical approach, but the method is easily extended to the panel case (without need for derivation of rate properties or variance formulae). In relaxing the strong stationarity asumption, the setup is specifically made to account for cases when treatment and covariates are not independent of time. For this reason, and to best explore treatment effect heterogeneity, we choose the Lu et al (2019) ATT estimator for our method. Their method also relies on a ``robust score",\footnote{With multiple nuisance functions in the framework, resulting in a ``quintuple-robust" score.} providing some of the efficiency improvements and bias reduction benefits of such a procedure, while being agnostic towards the specific choice of ML method used to obtain nuisance functions. 

Our proposed MLDID method extends this literature, building most closely on the approaches in \textcite{lu2019robust} and \textcite{callaway2021difference}.  We aim to estimate both group ATTs and predict unit level CATTs in robust way, incorporating ML in the estimation of nuisance models, allowing for the investigation of dynamic heterogeneity in the treatment effects.

\section{Methodology and Estimation Procedure}

\subsection{Staggered treatment adoption setup and potential outcomes}

We follow the notation by \textcite{callaway2021difference}.  We are interested in a setting with $T$ time periods, with particular times $t = 1,...,T$.    Let $D_{i,t}$  be a binary variable that indicates whether in a given time period $t$ a unit $i$ received treatment. The staggered treatment adoption setup assumes that for $t=2,...,T$, $D_{t-1}=1$ implies $D_t=1$,  with other words, no one is treated at time $t=1$ and once a unit becomes treated, they remain treated. 

We define $G$ as the time period when a unit first becomes treated, implying that for the units that eventually become treated, $G$ defines a group they belong to.  We further define a binary variable  $G_g$ that is equal to one if a unit first becomes treated in period $g$. 

Next we define potential outcomes in the staggered adoption setup, following  \textcite{callaway2021difference}.  Let $Y_{i,t}(0)$ denote a potential outcome for a unit  $i$ at time $t$ if they remain untreated though for the entire time period $T$.  For $g=2,...,T$, let  $Y_{i,t}(g)$ denote the potential outcome the same unit would experience at time $t$  if they were to start treatment in time period $g$. 

For a given unit only one of the potential outcomes is observed. The untreated potential outcome and the observed outcome $Y_{i,t}$ are related through  $Y_{i,t}=Y_{i,t}(0) + \sum_{g=2}^{T} (Y_{i,t}(g) - Y_{i,t}(0)) \cdot G_{i,g}$. In other words, the observed outcome is a sum of the treatment-free potential outcome and the treatment effect $Y_{i,t}(g) - Y_{i,t}(0)$ specific to the group where a given unit started treatment.

\subsection{The Group-Time Average Treatment Effect Parameter }

The causal-parameter of interest to us is the group-time average treatment effect on the treated, i.e., the average treatment effect  at a particular time $t$, among  those who started treatment at time period $g$:

\begin{equation}
    ATT_{g,t} = E[Y_t(g) - Y_t(0)| G_g = 1].
\end{equation}

This quantity allows the investigation of heterogeneity in treatment effects across groups and over time. For example, by  holding   $g$ fixed, and varying time $t$, one can scrutinize how treatment evolves over time for a group that started treatment at the same time. 
On the other hand, it can also be highlighted how the treatment effect, measured at a given time $t$, varies across different groups $g$. This quantity can be further aggregated into overall ``event study" ATTs, which show how the treatment effect evolves on average for all groups exposed to treatment as the length of exposure increases. 

The identification assumptions necessary to estimate $ATT_{g,t}$ (outlined in more details in \textcite{callaway2021difference}) are the following:

\begin{itemize}
\item Limited  or no treatment anticipation: potential outcomes under a given treatment start $g$ equal to the untreated potential outcomes, in time periods before the treatment start. 
\item Conditional parallel trends based on a not-yet treated group:  $E[Y_t(0) - Y_{t-1}(0)|X,  G_g = 1]-E[Y_t(0) - Y_{t-1}(0)|X, D_S=0,  G_g = 0]$, such that $t \geq g - \delta$ and $t+ \delta \leq  s < \bar{g} $, where $\bar{g}$ is the maximum period for treatment start in the sample. This assumption imposes parallel trends between group $g$ and groups that are not-yet treated by time $t+ \delta$. 

\item Overlap: a nonzero fraction of the population starts treatment at period $g$, and the generalized propensity score is bounded away from 1 (see more details in \textcite{callaway2021difference}).
\end{itemize}

\subsection{Estimation via Non-Parametric DiD}

\textcite{callaway2021difference} estimate the group-time ATT, $ATT_{g,t}$, in a series of two-period regressions.  We build on this, following the estimation technique proposed by \textcite{lu2019robust}, which  allows for incorporation of ML. Beyond the identifying assumptions already outlined,  this approach also relaxes the stationarity assumption and also allow the treatment effect to be heterogeneous according to observable characteristics $X$. 

We aim to estimate $ATT_{g,t}$  for all $g$ groups that are observed in the data, evaluated at all time periods between treatment start $g$  and $T$.  We do this by implementing a series of DiD estimators that contrast the observed outcomes of a group that started treatment at $g$, and a control group that consists of units that have not started treatment at period $g$ yet.   Each time the two-period data is subset, the post-treatment period will correspond to one of $g = g, g+1, ... T-g$, and the pre-treatment period will be held fixed as $g-1$. 
So, for example, to evaluate  $ATT_{2,3}$  we  use the baseline ($t=2-1$) and post-treatment ($t=3$) outcomes of the group that started treatment at time period $2$,  and the  baseline ($t=1$) and post-treatment ( $t=3$) outcomes of the group that has not yet started treatment at time period 2.

In the remainder of the section,  we slightly modify (and simplify) our notation, to reflect this two-period setup, closely following the notation of \textcite{lu2019robust}. We define a new binary variable $T_i$ that takes the value of $0$ for the pre-treatment observations and the value of $1$ for post-treatment observations.  $Y_i$ is the observed outcome of unit $i$ , $X_i$ is the vector of observed covariates, $G_{i,g}$ is a binary variable indicating whether a unit is in the treatment group (whether the unit has started treatment by time period $g$), and $T_i$ is an indicator variable for the second time period.\footnote{$Y_i$ does not have a time indicator as \textcite{lu2019robust} have developed their framework for a cross-sectional setup, which is equivalent to a two time period DiD setting.}

In each two-period subset, we assume the observed data is generated according to the process:
\begin{equation}
\label{eqn:outcomemodel}
    Y_i = \alpha(X_i) + \xi(X_i) \cdot G_{i,g} + \rho(X_i) \cdot T_i + \tau(X_i) \cdot G_{i,g}T_i + \epsilon_i,
\end{equation}

\noindent where $\xi(X_i)$ is the effect of the being in the treatment group on the expected outcome:

 \begin{equation}
  \begin{split}
     \xi(X_i) = & E[Y_i | X_i = x, G_i = 1, T_i = 0] - E[Y_i | X_i = x, G_i = 0, T_i = 0] \   
  \end{split} 
 \end{equation}
This can be thought of as the  difference  between parallel trends  in DiD,  while  here this difference is allowed to depend on X. The term $\rho(X_i)$ is the effect of time period on the expected outcome, again allowed to vary with $X$:
 \begin{equation}
  \begin{split}
     \rho(X_i) = & E[Y_i | X_i = x, G_i = 0, T_i = 1] - E[Y_i | X_i = x, G_i = 0, T_i = 0]
  \end{split} 
 \end{equation}

\noindent where $\tau(X_i)$ is the conditional average treatment effect, and the error term is assumed mean zero: $E[\epsilon_i| X_i, G_i, T_i] = 0$. Furthermore, we assume the joint distributions of $T_i, G_i$ and $X_i$ are arbitrary (i.e., they may not be independent). 

Because \textcite{lu2019robust} do not assume that treatment is independent of the joint distribution of time and $X_i$, the procedure for estimating nuisance functions is more complex than the recently proposed double-robust versions of \textcite{zimmert2020efficient} and \textcite{sant2020doubly}. In addition to a traditional propensity score (the probability of being treated given covariates $X_i$), we must also estimate the probability of being in each time period conditional on covariates $X_i$. We define these nuisance components as:

\begin{equation}
    g(x) = P[G_i = 1 | X_i = x], \;  t(x) = P[T_i = 1 | X_i = x].
\end{equation}

We also require an estimate for the joint conditional probability of treatment and time:

\begin{equation}
    \iota_{s,t}(x) = P[G_i = g, T_i = t | X_i = x],
\end{equation}

\noindent as well as a conditional response function (the outcome regression): 

\begin{equation}
    m(x) = E[Y_i | X_i = x].
\end{equation}

The nuisance components for the conditional response function marginalised over treatment and time, respectively, are:

\begin{equation}
\begin{split}
    \nu(x) = & E[Y_i | X_i = x, T_i = 1] - E[Y_i | X_i = x, T_i = 0], \\ 
    \zeta(x) = & E[Y_i | X_i = x, G_i = 1] - E[Y_i | X_i = x, G_i = 0],
\end{split}
\end{equation}

\noindent and the conditional covariance of $G_i$ and $T_i$ is
\begin{equation}
    \Delta(x) = \iota_{1,1} - g(x)t(x).
\end{equation}

As in \textcite{lu2019robust}, we estimate nuisance components using the R-learner of \textcite{nie2021quasi}. Note that the assumption in \textcite{sant2020doubly}, \textcite{zimmert2020efficient} and \textcite{callaway2021difference} that $G_i$ is independent of $T_i$ conditional on $X_i$ implies that $\Delta(X_i) = 0$. 

Given the outcome model (\ref{eqn:outcomemodel}), \textcite{lu2019robust} show that the data-generating distribution can be written as:
\begin{equation}
\label{eqn:outcomemodel2}
    Y_i = m(X_i) + A(X_i, G_i, T_i)\nu(X_i) + B(X_i, G_i, T_i)\zeta(X_i) + C(X_i, G_i, T_i)\tau(X_i) + \epsilon_i,
\end{equation}

\noindent where:
\begin{equation}
    \begin{split}
        A(\cdot) = & \left(1 - \frac{\Delta^2(X_i)}{g(X_i)(1-g(X_i))t(X_i)(1-t(X_i))}\right)^{-1}\left(T_i - t(X_i) - \frac{\Delta(X_i)(G_i - g(X_i)}{g(X_i)(1-g(X_i))}\right), \\
        B(\cdot) = & \left(1 - \frac{\Delta^2(X_i)}{g(X_i)(1-g(X_i))t(X_i)(1-t(X_i))}\right)^{-1}\left(S_i - s(X_i) - \frac{\Delta(X_i)(T_i - t(X_i)}{t(X_i)(1-t(X_i))}\right), \\
        C(\cdot) = & G_iT_i - \iota_{1,1}(X_i) - \left(g(X_i) + \frac{\Delta(X_i)}{t(X_i)}  \right)A(\cdot) - \left(t(X_i) + \frac{\Delta(X_i)}{g(X_i)}  \right)B(\cdot)
    \end{split}
\end{equation}

All terms in the above decomposition are orthogonal, and nuisance components of equation (\ref{eqn:outcomemodel2}) can be estimated using all of the data (as they are marginal quantities). 

\subsubsection{Heterogeneous Treatment Effects (CATT)}

In order to estimate the overall average treatment effect on the treated, we first need a good estimate of each $\tau_{g,t}(x)$:

\begin{equation}
    \begin{split}
        \tau_{g,t}(x) = & E[Y_i | X_i = x, G_i = 1, T_i = 1] - E[Y_i | X_i = x, G_i = 1, T_i = 0] \\
        & - E[Y_i | X_i = x, G_i = 0, T_i = 1] - E[Y_i | X_i = x, G_i = 0, T_i = 0].
    \end{split}
\end{equation}

\textcite{lu2019robust} propose estimating this function in the following way. First, we construct the quantity

\begin{equation}
    H(X_i, G_i, T_i) = Y_i - (m(X_i) + A(X_i, G_i, T_i)\nu(X_i) + B(X_i, G_i, T_i)\zeta(X_i).
\end{equation}

We then turn this into a loss function in order to estimate $\hat{\tau}(x)$:

\begin{equation}
    \hat{\tau}_{g,t}(x) = argmin_{\tau}\left\{\left(H(X_i, G_i, T_i) - C(X_i, G_i, T_i)\tau(X_i) \right)^2 + \Lambda\tau(X_i)\right\}
\end{equation}

We perform this step, as well as estimate all nuisance functions, using cross-validation techniques (described in the algorithm). 

\subsubsection{The Two-Period ATT}

After performing the previous steps, we are able to obtain our estimate of $E[\tau(x)]$, the ATT, as:

\begin{equation}
    ATT = \frac{1}{n}(\hat{\tau}(X_i) + \hat{\gamma}(X_i, G_i, T_i)(Y_i - \hat{y}(X_i, G_i, T_i))),
\end{equation}

\noindent where point estimates $\hat{y}(\cdot)$ are constructed as:
\begin{equation}
    \hat{y}(\cdot) = \hat{m}(X_i) + \hat{A}(X_i, G_i, T_i)\hat{\nu}(X_i) + \hat{B}(X_i, G_i, T_i)\hat{\zeta}(X_i) + \hat{C}(X_i, G_i, T_i)\hat{\tau}(X_i) 
\end{equation}

In other settings, the quantity $\hat{\gamma}$ usually takes the form of an inverse propensity score weighting scheme. In the DiD setting with $X_i$ and $G_i$ not independent from $T_i$, which requires the many forms of propensity scores described in Section 3.3, using a propensity score-based weighting scheme can be highly unstable, as the scores are much smaller.\footnote{This type of weighting is possible in our setting, but like in \textcite{lu2019robust}, we find that it leads to instability due to the very small propensity scores.} Therefore, \textcite{lu2019robust} suggest using the Augmented Minimax Linear Estimation (AMLE) approach of \textcite{hirshberg2021augmented}, which guarantees that $\gamma$ will not blow up when $e_{s,t}$ is small. AMLE directly estimates weights $\hat{\gamma}(\cdot)$ by solving a quadratic program which explicitly minimizes bias and variance:

\begin{equation}
\begin{split}
    \hat{\gamma}(\cdot) = & argmin_\gamma\left\{I^2_{\gamma, \mathcal{F}} + \frac{\sigma^2}{n^2}||\gamma||^2_2 : \gamma \in \mathcal{R}^n \right\}, \\
    I_{\gamma, \mathcal{F}} = & sup_{y\in\mathcal{F}}\left\{\frac{1}{n}\sum_{i=1}^n\left(\gamma(\cdot)y(\cdot) - (y(X_i,1,1) - y(X_i,0,1) - y(X_i,1,0) - y(X_i,0,0))
    \right)
    \right\}
  \end{split}  
\end{equation}

\noindent where $\mathcal{F}$ is some convex function class\footnote{We define this function class as in \textcite{lu2019robust}.} and $\sigma $ is an upper bound of $Var [ Y_i | X_i, G_i, T_i]$\footnote{This can be estimated using residuals, or as a tuning parameter in the optimization problem.}. 

\subsection{Dynamic Treatment Effects}

Now that we have obtained the group-time ATT for each two-period group, we can aggregate the estimates. In general, \textcite{callaway2021difference} uses the aggregation scheme:

\begin{equation}
    \theta = \sum_{g \in G} \sum_{t=2}^T w(g,t) ATT(g,t)
\end{equation}

The weighting function $w(g,t)$ can be chosen to highlight different forms of dynamic treatment effects (calendar or event time, for example). For an event-study specification, we wish to know how the average treatment effect varies according to length of exposure to treatment. In this case,

\begin{equation}
    w(g,t) = \mathbbm{1} \{g+e \leq T \}\mathbbm{1}\{t-g = e \}P(G = g | G + e \leq T)
\end{equation}

\noindent where $e$ is event time, $e=t-g$ is the time elapsed since treatment. Using these weights, ATTS are aggregated into their dynamic event study estimates (i.e., with respect to event time) by:

\begin{equation}
    \theta(e) = \sum_{g \in G} \mathbbm{1} \{g+e \leq T \} P(G = g | G + e \leq T) ATT(g,t).
\end{equation}

By extending this aggregation scheme of \textcite{callaway2021difference} to our CATT estimates, we also aggregate CATTs into their dynamic versions with respect to event time $e$:

\begin{equation}
    \theta_\tau(e) = \sum_{g \in G} \mathbbm{1} \{g+e \leq T \} P(G = g | G + e \leq T) \tau.
\end{equation}

\subsection{The ML-DiD Algorithm}

%Denote a subset of observed covariates as $Z \subseteq X$. For example, Z may be some poverty measure, or the Bolsa Familia coverage level in a municipality. We wish to estimate the conditional average treatment effect for these subgroups. Using the framework of Chernozhukov and Semenova (2017) as it is demonstrated in the DiD setting of Zimmert and Zimmert (2020), we will adapt the approach to a staggered treatment case by expanding $t = (0, 1, ..., T-1, T)$ (as opposed to $T = {0, 1}$).\footnote{This will require estimation of many propensity scores, i.e. a probability of being treated in each period.} To obtain the scores for this procedure, we will use the generalized version of a partial linear approach for machine learning estimation from Lu et al (2019) \parencite{lu2019robust,robinson1988root}. This yields conditional mean outcomes identical to Zimmert and Zimmert (2020) which will need to be reweighted by  $\frac{1}{Pr(D=1,T=t|Z)}$ (where treatment D is the introduction of FHP in a municipality). We will then linearly regress this weighted score on Z to obtain our desired estimates of treatment effect heterogeneity \parencite{chernozhukov2018simultaneous,zimmert2020paid}. To account for correlations across time, this will be done in a block bootstrap loop sampling procedure (where we resample observations across all time periods in which the observation is observed). 

In sum, the entire procedure is performed as follows: 

\begin{algorithm}
\caption{Algorithm for MLDiD}\label{alg:cap}
\begin{algorithmic}
\For{each group $g \in G$ }
\For{each time $t \in T$ }
\State $C \gets$ those not in $g$, not-yet-treated by time $t$ (control group)
\State Data $\gets$ subset where first period $= e-1$, and second period $ = t$ 
\State Obtain cross-fitted ML estimates of nuisance functions $g(x), \iota_{s,t}(x), m(x), \nu(x), \zeta(x)$
\State Compute $\delta(x), A(X_i,G_i,T_i), B(X_i,G_i,T_i), C(X_i,G_i,T_i), and H(X_i,G_i,T_i)$, and 
\[
\hat{y}(\cdot) = \hat{m}(X_i) + \hat{A}(X_i, G_i, T_i)\hat{\nu}(X_i) + \hat{B}(X_i, G_i, T_i)\hat{\zeta}(X_i) + \hat{C}(X_i, G_i, T_i)\hat{\tau}(X_i)
\]
\State Estimate $\hat{\tau}(x)$ using lasso, with loss function 
\[
\hat{\tau}(x) = argmin_{\tau}\left\{\left(H(X_i, G_i, T_i) - C(X_i, G_i, T_i)\tau(X_i) \right)^2 + \Lambda\tau(X_i)\right\}
\]
\State $\hat{\gamma} \gets$ AMLE as in Equation 16
\State Compute ATT for this group-time as
\[
ATT = \frac{1}{n}(\hat{\tau}(X_i) + \hat{\gamma}(X_i, G_i, T_i)(Y_i - \hat{y}(X_i, G_i, T_i))
\]
\EndFor
\EndFor
\State Aggregate ATT $\theta$, dynamic ATT $\theta(e)$ and dynamic CATTs $\theta_\tau(x)$ as in Eq 19 and 20
\end{algorithmic}
\end{algorithm}

\section{Simulations}

\subsection{Simulation Procedure}

For the first simulation design, we build off of the Monte Carlo procedure in \textcite{callaway2021difference}. The main difference of our version is that we allow the dynamic effect of participating in treatment to be heterogeneous. We generate four time-constant covariates, two continuous $X_1, X_2 \sim N(0,1)$ and two binary $X_3, X_4 \sim B(n, 0.5)$, and one time-varying covariate $X_5 \sim N(0,1) * t$. 

We set the probability of belonging to group $g \in G$ (i.e., the probability of being treated in a particular time period), to either random assignment (defined by $P(G=g|X) = \frac{1}{T+1}$) or by:\footnote{Note there is also a $\frac{1}{T+1}$ chance of being in the untreated group, $G=0$.}
\begin{equation}
    P(G=g|X) = \frac{exp((\kappa)'\gamma_g)}{\sum_{g\in G} exp((\kappa)'\gamma_g)}
\end{equation}

\noindent where $\kappa \in (X2, (X1 + X2 + X3))$ and $\gamma_g = 0.5 g/G$. Untreated potential outcomes are given by

\begin{equation}
Y_{it}(0) = \delta_t + \eta_i + u_{it}
\end{equation}

\noindent where $\delta_t = t$ (the time fixed effect), $\eta|G,X \sim N(G,1)$ is the individual fixed effect, and $u|G,X \sim N(0,1)$ is the error term. For treated units (those in group $g \neq 0$) in post-treatment periods ($t \geq g)$, we define the treated potential outcome as

%%%% NOTE THAT THIS HAS BEEN SIMPLIFIED FOR THIS DRAFT DUE TO THE CODE BUG AND WILL BE REMEDIED LATER %%%%%%%%%%%%%%%%%%%%% 
\begin{equation}
%Y_{it}(g) = \delta_t + \eta_i + \beta_t(\chi) + \delta_e\tau_i + v_{it}
Y_{it}(g) = \delta_e\tau_i + v_{it}
\end{equation}

%where $\beta_t \in (0, 1, t)$ (no effect, constant effect, calendar-time effect) and $\chi \in (X1, (X1 + X2 + X3)$. 

%In other words
%\begin{equation}
%Y_{it}(g) = Y_{it}(0) + \delta_e\tau_i + (v_{it} - u_{it})
%\end{equation}

\noindent where $\delta_e = e + 1 = t - g + 1$, $v|G,X,u_t \sim N(0,1)$, and the CATT $\delta_e\tau_i$ varies according to the scenarios

\begin{enumerate}
    \item $\tau(X_i) = X1$
    \item $\tau(X_i) = (X2 + X3)^2$
\end{enumerate}

Note that $e$ denotes the length of exposure to treatment, and the basic dynamic effect of treatment is simply equal to the number of periods a group has been treated. The conditional treatment effect heterogeneity is therefore introduced by $\tau_i$. Our setup differs from \textcite{callaway2021difference} by including $\tau$, as well as by expanding the number of covariates in $X_i$. We will repeat each scenario 100 times at sample sizes of $N =1000$, $N = 5000$, and $N = 10000$, and report performance in terms of root-mean-squared errors. We use number of periods $= 4$ (with 8 forthcoming).

For further examination of the performance of our estimator, we also modify the treated potential outcome as follows: 

\begin{equation}
Y_{it}(g) = \delta_t + \eta_i + \beta_t(\chi) + \delta_e\tau_i + v_{it}
\end{equation}

\noindent where $\beta_t \in (0, 1, t)$ (no effect, constant effect, calendar-time effect) and $\chi \in (X1, (X1 + X2 + X3)$. In other words,
\begin{equation}
Y_{it}(g) = Y_{it}(0) + \delta_e\tau_i + (v_{it} - u_{it}).
\end{equation}

Simulations for this extensive setting are ongoing, and preliminary investigation indicates results are consistent with initial findings. 

\subsection{Simulation Results: ATTs}

Table \ref{table:RMSE_gtts} compares RMSEs for group-time ATTs obtained by MLDID and DRDID (calculated as the difference between the oracle GTT and the estimated versions). Panel A depicts treatment effect heterogneiety $\tau = X_1$, and Panel B is $\tau = (X_2 + X_3)^2$. The left side is random treatment assignment, while the right side has propensity score determined by $X_2$. We see that although the DRDID of Callaway \& Sant'Anna achieves lower RMSEs than our proposed method, in settings where heterogeneity is more complex, and there is also more effect heterogeneity over time, they perform similarly. In Panel B, where the effect heterogeneity is more complex, the MLDID method achieves lower RMSEs than in Pane l A. Table \ref{table:simgroupATTs} depicts an example of the point estimates for each group-time ATT, compared to the oracle ATT, and shows that, even in a single repetition, our MLDID comes very close to both the oracle and the DRDID method.

\begin{table}[H]
\caption{RMSE, Simulated Group-Time ATTs}
\centering
\begin{tabular}{rrr|rr|rr||rr|rr|rr}
  \hline
  \multicolumn{4}{l}{\textbf{Panel A: $\tau = X_1$}} \\
  & \multicolumn{6}{c}{$\beta=0, \chi = 0$ (No Confounding)} & \multicolumn{6}{c}{$\kappa = X_1, \beta=1 \chi = X_1$}\\
  \hline
  & \multicolumn{2}{c}{N=2500} & \multicolumn{2}{c}{N=5000} & \multicolumn{2}{c}{N=10000} & \multicolumn{2}{c}{N=2500} & \multicolumn{2}{c}{N=5000} & \multicolumn{2}{c}{N=10000}\\
 & ML & DR & ML & DR & ML & DR  & ML & DR & ML & DR & ML & DR\\ 
  \hline
$ATT_{2,1}$ & 0.00 & 0.00 & 0.00 & 0.00 & 0.00 & 0.00 & 0.00 & 0.00 & 0.00 & 0.00 & 0.00 & 0.00 \\ 
  $ATT_{2,2}$ & 0.13 & 0.07 & 0.10 & 0.05 & 0.07 & 0.04 & 0.12 & 0.07 & 0.10 & 0.05 & 0.07 & 0.04 \\ 
  $ATT_{2,3}$ & 0.17 & 0.08 & 0.12 & 0.05 & 0.08 & 0.04 & 0.17 & 0.08 & 0.13 & 0.06 & 0.09 & 0.04 \\ 
  $ATT_{2,4}$ & 0.17 & 0.09 & 0.12 & 0.06 & 0.09 & 0.05 & 0.16 & 0.09 & 0.13 & 0.06 & 0.08 & 0.04 \\ 
  \hdashline
  $ATT_{3,1}$ & 0.11 & 0.08 & 0.07 & 0.05 & 0.05 & 0.04 & 0.11 & 0.08 & 0.07 & 0.05 & 0.06 & 0.04 \\ 
  $ATT_{3,2}$ & 0.00 & 0.00 & 0.00 & 0.00 & 0.00 & 0.00 & 0.00 & 0.00 & 0.00 & 0.00 & 0.00 & 0.00 \\ 
  $ATT_{3,3}$ & 0.41 & 0.08 & 0.81 & 0.06 & 0.28 & 0.04 & 1.14 & 0.08 & 0.35 & 0.06 & 0.71 & 0.04 \\ 
  $ATT_{3,4}$ & 0.16 & 0.09 & 0.11 & 0.06 & 0.08 & 0.04 & 0.16 & 0.10 & 0.12 & 0.06 & 0.08 & 0.04 \\ 
  \hdashline
  $ATT_{4,1}$ & 0.12 & 0.09 & 0.08 & 0.06 & 0.06 & 0.04 & 0.12 & 0.10 & 0.08 & 0.07 & 0.06 & 0.04 \\ 
  $ATT_{4,2}$ & 0.37 & 0.09 & 0.19 & 0.07 & 0.12 & 0.04 & 0.48 & 0.09 & 0.20 & 0.07 & 0.10 & 0.05 \\ 
  $ATT_{4,3}$ & 0.00 & 0.00 & 0.00 & 0.00 & 0.00 & 0.00 & 0.00 & 0.00 & 0.00 & 0.00 & 0.00 & 0.00 \\ 
  $ATT_{4,4}$ & 0.14 & 0.09 & 0.10 & 0.06 & 0.07 & 0.04 & 0.15 & 0.09 & 0.10 & 0.07 & 0.08 & 0.05 \\ 
   \hline
   \hline 
     \multicolumn{4}{l}{\textbf{Panel B: $\tau = (X_1 + X_3)^2$}} \\
  & \multicolumn{6}{c}{$\beta=0, \chi = 0$ (No Confounding)} & \multicolumn{6}{c}{$\kappa = X_1, \beta=1 \chi = X_1$}\\
  \hline
  & \multicolumn{2}{c}{N=2500} & \multicolumn{2}{c}{N=5000} & \multicolumn{2}{c}{N=10000} & \multicolumn{2}{c}{N=2500} & \multicolumn{2}{c}{N=5000} & \multicolumn{2}{c}{N=10000}\\
 & ML & DR & ML & DR & ML & DR  & ML & DR & ML & DR & ML & DR\\ 
  \hline
$ATT_{2,1}$ & 0.00 & 0.00 & 0.00 & 0.00 & 0.00 & 0.00 & 0.00 & 0.00 & 0.00 & 0.00 & 0.00 & 0.00 \\ 
  $ATT_{2,2}$ & 0.16 & 0.07 & 0.11 & 0.05 & 0.08 & 0.04 & 0.17 & 0.07 & 0.14 & 0.05 & 0.12 & 0.04 \\ 
  $ATT_{2,3}$ & 0.23 & 0.08 & 0.15 & 0.05 & 0.12 & 0.04 & 0.21 & 0.08 & 0.16 & 0.06 & 0.11 & 0.04 \\ 
  $ATT_{2,4}$ & 0.18 & 0.09 & 0.13 & 0.06 & 0.09 & 0.05 & 0.53 & 0.09 & 0.51 & 0.06 & 0.48 & 0.04 \\ 
  \hdashline
  $ATT_{3,1}$ & 0.11 & 0.08 & 0.07 & 0.05 & 0.05 & 0.04 & 0.11 & 0.08 & 0.07 & 0.05 & 0.06 & 0.04 \\ 
  $ATT_{3,2}$ & 0.00 & 0.00 & 0.00 & 0.00 & 0.00 & 0.00 & 0.00 & 0.00 & 0.00 & 0.00 & 0.00 & 0.00 \\ 
  $ATT_{3,3}$ & 0.36 & 0.08 & 0.31 & 0.06 & 0.14 & 0.04 & 0.55 & 0.08 & 0.37 & 0.06 & 0.31 & 0.04 \\ 
  $ATT_{3,4}$ & 0.21 & 0.09 & 0.15 & 0.06 & 0.11 & 0.04 & 0.79 & 0.10 & 0.74 & 0.06 & 0.73 & 0.04 \\ 
  \hdashline
  $ATT_{4,1}$ & 0.12 & 0.09 & 0.08 & 0.06 & 0.06 & 0.04 & 0.12 & 0.10 & 0.08 & 0.07 & 0.06 & 0.04 \\ 
  $ATT_{4,2}$ & 0.37 & 0.09 & 0.19 & 0.07 & 0.12 & 0.04 & 0.48 & 0.09 & 0.20 & 0.07 & 0.10 & 0.05 \\ 
  $ATT_{4,3}$ & 0.00 & 0.00 & 0.00 & 0.00 & 0.00 & 0.00 & 0.00 & 0.00 & 0.00 & 0.00 & 0.00 & 0.00 \\ 
  $ATT_{4,4}$ & 0.17 & 0.09 & 0.11 & 0.06 & 0.08 & 0.04 & 0.76 & 0.09 & 0.75 & 0.07 & 0.76 & 0.05 \\ 
   \hline
\end{tabular}
\label{table:RMSE_gtts}
{\footnotesize \justifying \singlespacing{This table presents RMSEs of group-time ATT estimates $ATT_{g,t}$ for each treatment group $g$ and $t$ pair. The first column is our proposed estimator and the second column is estimates from the \textcite{callaway2021difference} DRDID method. We report results for 500 repetitions of the simulations. } \par}
\end{table}

\begin{table}
\centering
    \begin{threeparttable}
\caption{Simulation Results: Group-Time ATTs}
\label{table:simgroupATTs}
\begin{tabular}{l c c c | c c c }
\hline
\multicolumn{3}{l}{\textbf{Random TX, $\beta = 1$}} & & \multicolumn{3}{l}{\textbf{$P(G|X=X_2)$, $\beta = 1$}} \\
& \multicolumn{3}{c}{N=5000} \\
 & ORACLE & MLDID & DRDID & ORACLE & MLDID & DRDID\\
\hline
$ATT_{2,1}$   & $-$   & $-$   & $-$ & $-$   & $-$   & $-$  \\

$ATT_{2,2}$   & $-4.08$  & $-4.03$  & $-4.06$  & $-4.02$  & $-4.00$  & $-4.00$ \\
     & $(0.00)$ & $(0.07)$ & $(0.10)$ & $(0.00)$ & $(0.07)$ & $(0.11)$ \\
$ATT_{2,3}$   & $-5.00$  & $-5.02$  & $-5.00$ & $-5.03$  & $-4.97$  & $-5.02$ \\
     & $(0.00)$ & $(0.08)$ & $(0.14)$ & $(0.00)$ & $(0.08)$ & $(0.12)$ \\
$ATT_{2,4}$   & $-6.10$  & $-6.07$  & $-6.10$ & $-5.95$  & $-5.87$  & $-5.97$ \\
     & $(0.00)$ & $(0.10)$ & $(0.17)$ & $(0.00)$ & $(0.10)$ & $(0.16)$ \\
\hdashline
$ATT_{3,1}$ & $0.00$   & $0.07$   & $0.05$  & $0.00$   & $0.05$   & $0.06$ \\
     & $(0.00)$ & $(0.09)$ & $(0.05)$ & $(0.00)$ & $(0.09)$ & $(0.06)$ \\
$ATT_{3,2}$ & $-$   & $-$   & $-$ & $-$   & $-$   & $-$  \\
$ATT_{3,3}$ & $-4.94$  & $-4.98$  & $-4.90$ & $-5.01$  & $-4.96$  & $-4.94$ \\
     & $(0.00)$ & $(0.08)$ & $(0.10)$ & $(0.00)$ & $(0.08)$ & $(0.10)$ \\
$ATT_{3,4}$ & $-5.75$  & $-5.92$  & $-5.86$ & $-5.90$  & $-5.85$  & $-5.88$ \\
     & $(0.00)$ & $(0.09)$ & $(0.12)$ & $(0.00)$ & $(0.09)$ & $(0.13)$\\
\hdashline
$ATT_{4,1}$ & $0.00$   & $0.05$   & $0.05$ & $0.00$   & $0.11$   & $0.15$  \\
     & $(0.00)$ & $(0.09)$ & $(0.06)$ & $(0.00)$ & $(0.09)$ & $(0.07)$\\
$ATT_{4,2}$ & $0.00$   & $0.06$   & $0.08$ & $0.00$   & $0.05$   & $0.08$  \\
     & $(0.00)$ & $(0.09)$ & $(0.06)$ & $(0.00)$ & $(0.09)$ & $(0.06)$\\
$ATT_{4,3}$ & $-$   & $-$   & $-$ & $-$   & $-$   & $-$  \\
$ATT_{4,4}$ & $-6.24$  & $-6.12$  & $-6.14$ & $-6.08$  & $-5.93$  & $-6.03$  \\
     & $(0.00)$ & $(0.08)$ & $(0.10)$ & $(0.00)$ & $(0.08)$ & $(0.11)$ \\
\hline 
\end{tabular}
\begin{tablenotes}
            \footnotesize
            \item[]\justifying\singlespacing
            This table presents group-time ATT estimates $ATT_{g,t}$ for each treatment group $g$ and $t$ pair. The first column is our proposed estimator and the second column is estimates from the \textcite{callaway2021difference} DRDID method. The Oracle column is the true simulated group-time ATT. The sample size is $N=5000$ and we report results for a single repetition (other sample sizes and iterations forthcoming).
        \end{tablenotes}
    \end{threeparttable}
\end{table}

We also ensured that our aggregation procedure (which transforms the group-time ATTS into their dynamic, event-study versions) was correct by examining RMSE and raw estimates of the dynamic ATTs. As an example, we show two versions of the simulations in Table \ref{table:simdynamicATTs}, demonstrating that even a single small repetition of the simulation gets very close to the oracle and the DRDID version. %Table \ref{table:RMSE_dynatts} shows the same results in terms of RMSE. 

\begin{table}[ht]
\centering
\begin{threeparttable}
\caption{Simulation Results: Dynamic ATTs}
\label{table:simdynamicATTs}
\begin{tabular}{rccc | ccc }
  \hline
  \multicolumn{3}{l}{\textbf{Random TX, dynamic het}} & & \multicolumn{3}{l}{\textbf{$P(G|X=X_2)$, dynamic het}}\\
  & \multicolumn{3}{c}{N=5000} \\
 Event-time & ORACLE & MLDID & DRDID & ORACLE & MLDID & DRDID \\ 
  \hline
-3 & $0.00$   & $0.05$   & $0.05$ & $0.00$   & $0.11$   & $0.15$  \\
   & $(0.00)$ & $(0.09)$ & $(0.06)$ & $(0.00)$ & $(0.09)$ & $(0.06)$ \\
-2 & $0.00$   & $0.07$   & $0.07$  & $0.00$   & $0.05$   & $0.07$ \\
   & $(0.00)$ & $(0.09)$ & $(0.04)$ & $(0.00)$ & $(0.09)$ & $(0.04)$\\
-1 & $-$   & $-$   & $-$ & $-$   & $-$   & $-$  \\
0  & $-5.08$  & $-5.04$  & $-5.07$ & $-5.04$  & $-4.96$  & $-5.01$ \\
   & $(0.00)$ & $(0.08)$ & $(0.06)$ & $(0.00)$ & $(0.08)$ & $(0.06)$\\
1  & $-5.38$  & $-5.47$  & $-5.44$ & $-5.46$  & $-5.41$  & $-5.46$ \\
   & $(0.00)$ & $(0.09)$ & $(0.09)$ & $(0.00)$ & $(0.09)$ & $(0.09)$\\
2  & $-6.10$  & $-6.07$  & $-6.10$ & $-5.95$  & $-5.87$  & $-5.97$ \\
   & $(0.00)$ & $(0.10)$ & $(0.17)$ & $(0.00)$ & $(0.10)$ & $(0.15)$\\
   \hline
\end{tabular}
\begin{tablenotes}
            \footnotesize
            \item[]\justifying\singlespacing
            This table presents dynamic ATT estimates. The first column is our proposed estimator and the second column is estimates from the \textcite{callaway2021difference} DRDID method. The Oracle column is the true simulated dynamic ATT. The sample size is $N=5000$ and we report results for a single repetition (other sample sizes and iterations available on request).
        \end{tablenotes}
    \end{threeparttable}
%\label{table:simdynamicATTs}
\end{table}

%% ADD DYNAMIC RMSE HERE (GRAPHS) 

\subsection{Simulation Results: CATTs}

Table \ref{table:RMSE_cates} presents RMSE results for the estimated dynamic CATTs and scores, as compared to the oracle CATTs. The estimted CATTs perform better than estimated DR-scores in all cases. We note that there is not much improvement as sample size increases, and this is a point of ongoing investigation. 

\begin{table}[H]
\caption{RMSE, Dynamic CATTs and DR-scores}
\centering
\begin{tabular}{rrr|rr|rr||rr|rr|rr}
  \hline
  \multicolumn{4}{l}{\textbf{Panel A: $\tau = X_1$}} \\
  & \multicolumn{6}{c}{$\beta=0, \chi = 0$ (No Confounding)} & \multicolumn{6}{c}{$\kappa = X_1, \beta=1 \chi = X_1$}\\
  \hline
  & \multicolumn{2}{c}{N=2500} & \multicolumn{2}{c}{N=5000} & \multicolumn{2}{c}{N=10000} & \multicolumn{2}{c}{N=2500} & \multicolumn{2}{c}{N=5000} & \multicolumn{2}{c}{N=10000}\\
 $e$& $\hat{\tau}$ & $\hat{\Gamma}$ & $\hat{\tau}$ & $\hat{\Gamma}$ & $\hat{\tau}$ & $\hat{\Gamma}$ & $\hat{\tau}$ & $\hat{\Gamma}$ & $\hat{\tau}$ & $\hat{\Gamma}$ & $\hat{\tau}$ & $\hat{\Gamma}$\\ 
  \hline
-2 & 1.94 & 3.27 & 1.94 & 3.28 & 1.94 & 3.28 & 1.96 & 3.33 & 1.95 & 3.33 & 1.95 & 3.34 \\ 
  -1 & 1.56 & 2.89 & 1.55 & 2.89 & 1.52 & 2.88 & 1.58 & 2.95 & 1.56 & 2.95 & 1.53 & 2.89 \\ 
  0 & 0.00 & 0.00 & 0.00 & 0.00 & 0.00 & 0.00 & 0.00 & 0.00 & 0.00 & 0.00 & 0.00 & 0.00 \\ 
  1 & 3.30 & 3.99 & 3.29 & 3.92 & 3.30 & 3.94 & 3.34 & 4.02 & 3.32 & 3.99 & 3.34 & 4.00 \\ 
  2 & 3.50 & 3.90 & 3.52 & 3.93 & 3.53 & 3.94 & 3.52 & 3.94 & 3.53 & 3.97 & 3.55 & 3.98 \\ 
  3 & 3.73 & 3.98 & 3.74 & 4.00 & 3.76 & 4.01 & 3.74 & 4.03 & 3.76 & 4.05 & 3.77 & 4.06 \\ 
   \hline
   \hline 
     \multicolumn{4}{l}{\textbf{Panel B: $\tau = (X_1 + X_3)^2$}} \\
  & \multicolumn{6}{c}{$\beta=0, \chi = 0$ (No Confounding)} & \multicolumn{6}{c}{$\kappa = X_1, \beta=1 \chi = X_1$}\\
  \hline
  & \multicolumn{2}{c}{N=2500} & \multicolumn{2}{c}{N=5000} & \multicolumn{2}{c}{N=10000} & \multicolumn{2}{c}{N=2500} & \multicolumn{2}{c}{N=5000} & \multicolumn{2}{c}{N=10000}\\
 $e$& $\hat{\tau}$ & $\hat{\Gamma}$ & $\hat{\tau}$ & $\hat{\Gamma}$ & $\hat{\tau}$ & $\hat{\Gamma}$ & $\hat{\tau}$ & $\hat{\Gamma}$ & $\hat{\tau}$ & $\hat{\Gamma}$ & $\hat{\tau}$ & $\hat{\Gamma}$\\ 
  \hline
-2 & 1.70 & 3.13 & 1.70 & 3.14 & 1.70 & 3.14 & 1.76 & 3.22 & 1.75 & 3.22 & 1.75 & 3.23 \\ 
  -1 & 1.34 & 2.79 & 1.32 & 2.79 & 1.30 & 2.79 & 1.36 & 2.85 & 1.34 & 2.86 & 1.31 & 2.79 \\ 
  0 & 0.00 & 0.00 & 0.00 & 0.00 & 0.00 & 0.00 & 0.00 & 0.00 & 0.00 & 0.00 & 0.00 & 0.00 \\ 
  1 & 2.11 & 4.02 & 2.11 & 4.00 & 2.13 & 4.01 & 2.25 & 4.12 & 2.25 & 4.15 & 2.25 & 4.15 \\ 
  2 & 2.73 & 4.48 & 2.75 & 4.50 & 2.75 & 4.50 & 2.77 & 4.67 & 2.78 & 4.71 & 2.78 & 4.73 \\ 
  3 & 4.22 & 4.50 & 4.23 & 4.49 & 4.24 & 4.47 & 3.96 & 4.78 & 3.97 & 4.84 & 3.98 & 4.87 \\ 
   \hline
\end{tabular}
\label{table:RMSE_cates}
{\footnotesize \justifying \singlespacing{This table presents RMSEs of CATT estimates and DR score estimates, compared to the oracles. The first column is the estimated CATTs and the second is the DR scores. } \par}
\end{table}

\subsection{Simulation Results: BLPs}

We compare the coefficients from regressing the oracle CATTs onto covariates with the coefficients from regression estimated CATTs and DR scores on covariates. That is, we perform the following estimations : 

\begin{equation}
    \hat{\tau_{i,e}} = \alpha_{i,t} + \beta_{\tau}(X_{i,e}) + \zeta_e + \epsilon 
\end{equation}
and 
\begin{equation}
    \hat{\Gamma_{i,e}} = \alpha_{i,t} + \beta_{\Gamma}(X_{i,e}) + \zeta_e + \epsilon
\end{equation}

\noindent where $\Gamma = \hat{\tau}(X_i) + \hat{\gamma}(X_i, G_i, T_i)(Y_i - \hat{y}(X_i, G_i, T_i))$ is the robust score (akin to the doubly-robust score in other settings), $\zeta_t$ is a time fixed-effect (indicator for each event time). The covariates that are associated with heterogeneity are picked up as significant by the regressions and increase correctly over time (as designed in the DGP). Figure \ref{BLP_coefsims} depicts the coefficients from the first simulation scenario, where the effect heterogeneity is only influenced by $X1$. As expected, the coefficients only pick up on this covariate (as opposed to the covariates on $X2$, which show no significant heterogeneity), and the magnitude is very close to the oracle. Table \ref{tab:avgBLPs} shows average coefficients from the simulated BLPs. In exploratory work (forthcoming), we find that when confounding is present, the CATT estimates may erroneously pick up a significant effect of the confounding covariate, although the magnitude remains small. For this reason, we recommend making use of the DR scores in the BLP analysis for real-world settings. 

\begin{figure}[H]
    \centering
    \caption{BLP of Heterogeneity}
    \addtocounter{figure}{-1}
\begin{subfigure}{0.48\linewidth}
  \centering
  \includegraphics[width=\linewidth]{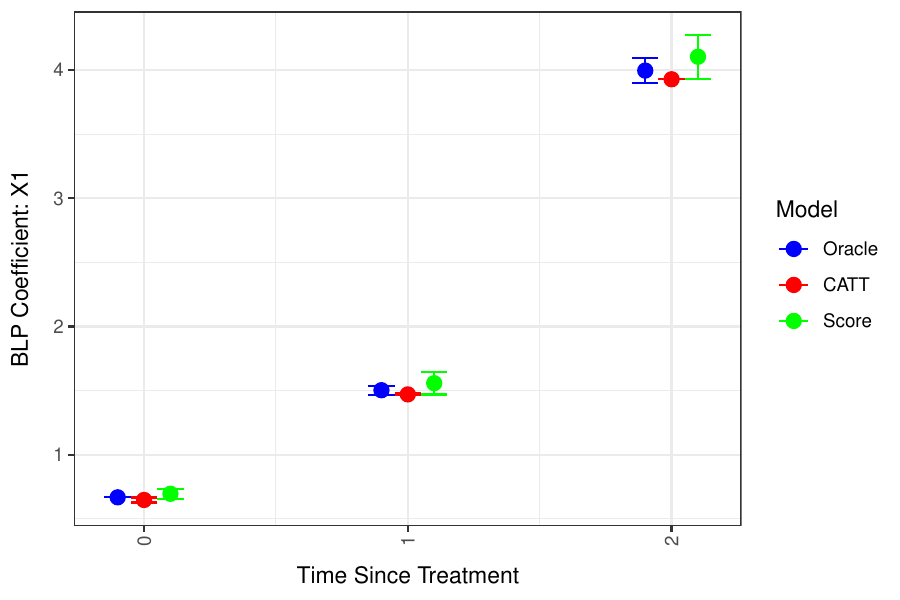}
  \caption{Heterogeneity in X1}
  %\label{fig: Image1} 
\end{subfigure}
\quad
\begin{subfigure}{0.48\linewidth}
  \centering
  \includegraphics[width=\linewidth]{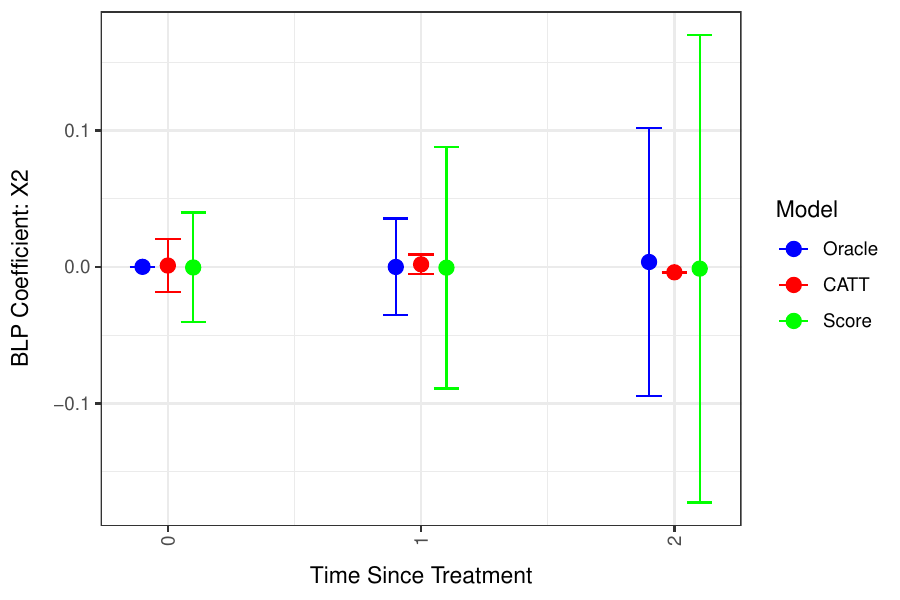}
  \caption{Heterogeneity in X2}
  %\label{fig:NDRvaluecalc} 
\end{subfigure}
%{\footnotesize \justifying \singlespacing{This figure depicts overlap scenarios for $\psi \in (1,2)$ } \par}
\caption*{\footnotesize{This figure depicts simulation results for a BLP regression where the true treatment effect is only determined by $X1$. The left panel depicts coefficients from this covariate and the right panel depicts another covariate which does not have an effect on $\tau$.}}
\label{BLP_coefsims}
\end{figure}

\begin{table}[ht]
\caption{AVG BLP coefficient by event-time (n=2500, 100 repetitions)}
\centering
\begin{tabular}{rrrr|rrr|rrr}
  \hline
 & \multicolumn{3}{c}{e=0} & \multicolumn{3}{c}{e=1} & \multicolumn{3}{c}{e=2} \\
 & OR & CATTs & scores & OR & CATTs & scores & OR & CATTs & scores \\ 
  \hline
(Intercept) & -1.63 & -3.20 & -3.68 & -2.91 & -2.86 & -2.84 & -6.01 & -5.93 & -5.97 \\ 
  \textbf{X1} & \textbf{0.64} & \textbf{0.62} & \textbf{0.69} & \textbf{1.52} & \textbf{1.47} & \textbf{1.69} & \textbf{3.88} & \textbf{3.73} & \textbf{3.96} \\

  X2 & -0.05 & -0.26 & -0.28 & -0.02 & -0.05 & -0.07 & -0.04 & 0.00 & -0.27 \\ 
  X3 & 0.01 & 0.13 & 0.39 & 0.13 & 0.10 & 0.27 & 0.08 & -0.00 & 0.19 \\ 
  X4 & -0.02 & 0.11 & 0.30 & 0.05 & 0.04 & 0.16 & -0.03 & -0.00 & -0.01 \\ 
  X5 & 0.06 & 0.30 & 0.38 & -0.05 & -0.11 & 0.10 & 0.08 & 0.00 & -0.15 \\ 
   \hline
\end{tabular}
\caption*{\footnotesize{This table shows average coefficients from regression of the estimated CATTs and robust scores on the simulated covaraites, compared to regressing the oracle CATTs on the covariates. The bold row is significant at the 1\% level and corresponds to the true covariate that impacts treatment effect heterogeneity, i.e., $\tau = X_1$. Each of the three panels corresponds to an event-time, so $e=1$ corresponds to one period since treatment}}
\label{tab:avgBLPs}
\end{table}

\subsection{Simulation Results: Classification Analysis (CLAN)}

There may be settings where the researcher wishes to examine univariate treatment effect heterogeneity. That is, we would like to know whether there is a difference according to individual covariates for the most and least affected groups, and to examine whether the difference evolves over time, without taking other covariates into account. Following \textcite{chernozhukov2018generic}, we focus on the least and most affected subgroups, $S_1$ and $S_K$. The average characteristics of the least and most affected groups are estimated as:

\begin{equation}
    \hat{\delta}_i = E[s(X_i, Y_i) | S_i],
\end{equation}

\noindent where $i = 1$ is the least and $i = K$ the most affected groups (i.e., those with the lowest and highest CATTs) and $s(X_i, Y_i)$ is a vector of unit characteristics. We can then use standard techniques to test the difference between these estimates. 

Table \ref{table:clans} shows the average results for the simulation CLANs for the N=10,000 settings, separated by event time. We see that the difference $\hat{\delta}_K - \hat{\delta}_1$ is significant and in the correct direction for those covariates which are indeed driving the treatment effect heterogeneiety ($X_1$ for Panel A and B, and $X_2, X_3$ for Panel C and D), although the magnitude is not always consistent with the oracle. We also note that $X_5$ is picked up as having a significant difference in most and least affected groups for the first event time, which bears further investigation. As a whole, we recommend use of the CLAN to confirm findings of the BLP in terms of which variables have been picked up as significant, and to examine the direction of effect. 

\begin{table}
\caption{Simulation Results: CLANs (N=10,000)}
\begin{center}
\begin{adjustbox}{width=\textwidth}
\begin{tabular}{l c c c| c c c | c c c}
\hline
\multicolumn{7}{l}{\textbf{Panel A: $\tau = X_1$, No Confounding}} \\
& \multicolumn{3}{c}{e=0} &\multicolumn{3}{c}{e=1} & \multicolumn{3}{c}{e=2} \\
 & OR & CATT & SCORE & OR & CATT & SCORE & OR & CATT & SCORE \\
\hline
X1 & $1.80^{*}$       & $1.93^{*}$       & $1.15^{*}$       & $2.22^{*}$        & $2.76^{*}$        & $1.42^{*}$       & $2.44^{*}$       & $2.80^{*}$       & $2.04^{*}$       \\
   & $ [ 1.70; 1.89]$ & $ [ 1.87; 1.99]$ & $ [ 1.08; 1.23]$ & $ [ 2.12;  2.32]$ & $ [ 2.71;  2.80]$ & $ [ 1.32; 1.51]$ & $ [ 2.31; 2.57]$ & $ [ 2.73; 2.86]$ & $ [ 1.93; 2.15]$ \\
X2 & $-0.00$          & $0.00$           & $0.00$           & $-0.00$           & $0.01$            & $0.00$           & $-0.00$          & $0.00$           & $-0.00$          \\
   & $ [-0.12; 0.12]$ & $ [-0.08; 0.08]$ & $ [-0.08; 0.08]$ & $ [-0.14;  0.14]$ & $ [-0.09;  0.10]$ & $ [-0.10; 0.10]$ & $ [-0.21; 0.20]$ & $ [-0.14; 0.14]$ & $ [-0.14; 0.14]$ \\
X3 & $0.00$           & $0.00$           & $0.04$           & $0.00$            & $0.00$            & $-0.00$          & $0.00$           & $0.00$           & $-0.00$          \\
   & $ [-0.06; 0.06]$ & $ [-0.04; 0.04]$ & $ [-0.00; 0.08]$ & $ [-0.07;  0.07]$ & $ [-0.05;  0.05]$ & $ [-0.05; 0.04]$ & $ [-0.10; 0.10]$ & $ [-0.07; 0.07]$ & $ [-0.07; 0.07]$ \\
X4 & $-0.00$          & $0.00$           & $0.04$           & $0.00$            & $0.00$            & $-0.00$          & $0.00$           & $0.00$           & $-0.00$          \\
   & $ [-0.06; 0.06]$ & $ [-0.04; 0.04]$ & $ [-0.00; 0.08]$ & $ [-0.07;  0.07]$ & $ [-0.05;  0.05]$ & $ [-0.05; 0.05]$ & $ [-0.10; 0.10]$ & $ [-0.07; 0.07]$ & $ [-0.07; 0.07]$ \\
X5 & $0.26^{*}$       & $0.67^{*}$       & $0.49^{*}$       & $-0.16^{*}$       & $-0.20^{*}$       & $-0.07$          & $0.00$           & $-0.00$          & $-0.01$          \\
   & $ [ 0.12; 0.40]$ & $ [ 0.58; 0.75]$ & $ [ 0.39; 0.58]$ & $ [-0.32; -0.00]$ & $ [-0.31; -0.10]$ & $ [-0.17; 0.04]$ & $ [-0.20; 0.21]$ & $ [-0.14; 0.14]$ & $ [-0.14; 0.13]$ \\
\hline
\multicolumn{7}{l}{\textbf{Panel B: $\tau = X_1$, Confounding}} \\
& \multicolumn{3}{c}{e=0} &\multicolumn{3}{c}{e=1} & \multicolumn{3}{c}{e=2} \\
& OR & CATE & SCORE & OR & CATE & SCORE & OR & CATE & SCORE \\
\hline
X1 & $1.79^{*}$       & $1.92^{*}$        & $1.14^{*}$        & $2.22^{*}$        & $2.76^{*}$        & $1.39^{*}$       & $2.44^{*}$       & $2.80^{*}$       & $2.00^{*}$       \\
   & $ [ 1.70; 1.89]$ & $ [ 1.86;  1.98]$ & $ [ 1.07;  1.22]$ & $ [ 2.12;  2.32]$ & $ [ 2.71;  2.81]$ & $ [ 1.29; 1.48]$ & $ [ 2.31; 2.57]$ & $ [ 2.73; 2.86]$ & $ [ 1.89; 2.11]$ \\
X2 & $-0.07$          & $-0.19^{*}$       & $-0.11^{*}$       & $-0.02$           & $-0.03$           & $-0.02$          & $0.00$           & $-0.00$          & $-0.01$          \\
   & $ [-0.19; 0.04]$ & $ [-0.27; -0.11]$ & $ [-0.19; -0.03]$ & $ [-0.17;  0.12]$ & $ [-0.13;  0.07]$ & $ [-0.12; 0.08]$ & $ [-0.20; 0.20]$ & $ [-0.14; 0.14]$ & $ [-0.14; 0.13]$ \\
X3 & $-0.00$          & $-0.00$           & $0.03$            & $-0.00$           & $-0.00$           & $-0.00$          & $-0.00$          & $-0.00$          & $-0.00$          \\
   & $ [-0.06; 0.06]$ & $ [-0.04;  0.04]$ & $ [-0.01;  0.07]$ & $ [-0.07;  0.07]$ & $ [-0.05;  0.05]$ & $ [-0.05; 0.04]$ & $ [-0.11; 0.10]$ & $ [-0.07; 0.07]$ & $ [-0.07; 0.07]$ \\
X4 & $-0.00$          & $-0.00$           & $0.03$            & $-0.00$           & $-0.00$           & $-0.01$          & $-0.00$          & $-0.00$          & $-0.00$          \\
   & $ [-0.06; 0.06]$ & $ [-0.04;  0.04]$ & $ [-0.01;  0.07]$ & $ [-0.07;  0.07]$ & $ [-0.05;  0.05]$ & $ [-0.06; 0.04]$ & $ [-0.10; 0.10]$ & $ [-0.07; 0.07]$ & $ [-0.07; 0.07]$ \\
X5 & $0.26^{*}$       & $0.65^{*}$        & $0.48^{*}$        & $-0.16^{*}$       & $-0.20^{*}$       & $-0.07$          & $0.00$           & $0.00$           & $0.00$           \\
   & $ [ 0.12; 0.40]$ & $ [ 0.56;  0.74]$ & $ [ 0.39;  0.57]$ & $ [-0.32; -0.00]$ & $ [-0.31; -0.09]$ & $ [-0.17; 0.04]$ & $ [-0.21; 0.21]$ & $ [-0.14; 0.14]$ & $ [-0.14; 0.14]$ \\
\hline
\hline
\multicolumn{7}{l}{\textbf{Panel C: $\tau = (X_2+X_3)^2$, No Confounding}} \\
& \multicolumn{3}{c}{e=0} &\multicolumn{3}{c}{e=1} & \multicolumn{3}{c}{e=2} \\
 & OR & CATE & SCORE & OR & CATE & SCORE & OR & CATE & SCORE \\
\hline
X1 & $-0.00$          & $-0.00$          & $0.00$           & $0.00$           & $-0.00$           & $0.00$            & $-0.00$          & $-0.00$          & $0.00$           \\
   & $ [-0.12; 0.12]$ & $ [-0.08; 0.08]$ & $ [-0.08; 0.08]$ & $ [-0.14; 0.15]$ & $ [-0.10;  0.09]$ & $ [-0.10;  0.10]$ & $ [-0.21; 0.20]$ & $ [-0.14; 0.14]$ & $ [-0.14; 0.14]$ \\
X2 & $1.01^{*}$       & $2.22^{*}$       & $0.73^{*}$       & $1.09^{*}$       & $2.48^{*}$        & $0.52^{*}$        & $1.12^{*}$       & $2.51^{*}$       & $0.62^{*}$       \\
   & $ [ 0.89; 1.14]$ & $ [ 2.17; 2.27]$ & $ [ 0.64; 0.82]$ & $ [ 0.94; 1.24]$ & $ [ 2.42;  2.54]$ & $ [ 0.40;  0.63]$ & $ [ 0.91; 1.33]$ & $ [ 2.43; 2.59]$ & $ [ 0.46; 0.78]$ \\
X3 & $0.28^{*}$       & $0.53^{*}$       & $0.23^{*}$       & $0.30^{*}$       & $0.60^{*}$        & $0.13^{*}$        & $0.31^{*}$       & $0.61^{*}$       & $0.16^{*}$       \\
   & $ [ 0.22; 0.33]$ & $ [ 0.50; 0.57]$ & $ [ 0.19; 0.27]$ & $ [ 0.23; 0.37]$ & $ [ 0.56;  0.64]$ & $ [ 0.08;  0.18]$ & $ [ 0.22; 0.41]$ & $ [ 0.56; 0.67]$ & $ [ 0.09; 0.23]$ \\
X4 & $-0.00$          & $-0.00$          & $0.03$           & $0.00$           & $-0.00$           & $-0.02$           & $-0.00$          & $-0.00$          & $-0.02$          \\
   & $ [-0.06; 0.06]$ & $ [-0.04; 0.04]$ & $ [-0.01; 0.07]$ & $ [-0.07; 0.07]$ & $ [-0.05;  0.05]$ & $ [-0.07;  0.03]$ & $ [-0.10; 0.10]$ & $ [-0.07; 0.07]$ & $ [-0.09; 0.05]$ \\
X5 & $0.20^{*}$       & $0.29^{*}$       & $0.14^{*}$       & $-0.14$          & $-0.19^{*}$       & $-0.34^{*}$       & $-0.00$          & $0.00$           & $-0.00$          \\
   & $ [ 0.06; 0.35]$ & $ [ 0.20; 0.38]$ & $ [ 0.05; 0.23]$ & $ [-0.29; 0.02]$ & $ [-0.30; -0.09]$ & $ [-0.44; -0.23]$ & $ [-0.21; 0.20]$ & $ [-0.14; 0.14]$ & $ [-0.14; 0.13]$ \\
\hline
\multicolumn{7}{l}{\textbf{Panel D: $\tau = (X_2+X_3)^2$, Confounding}} \\
& \multicolumn{3}{c}{e=0} &\multicolumn{3}{c}{e=1} & \multicolumn{3}{c}{e=2} \\
& OR & CATE & SCORE & OR & CATE & SCORE & OR & CATE & SCORE \\
\hline
X1 & $-0.00$          & $-0.00$          & $-0.00$          & $0.00$           & $-0.01$           & $-0.00$           & $-0.00$          & $0.00$           & $0.00$           \\
   & $ [-0.12; 0.12]$ & $ [-0.08; 0.08]$ & $ [-0.08; 0.08]$ & $ [-0.14; 0.15]$ & $ [-0.10;  0.09]$ & $ [-0.10;  0.10]$ & $ [-0.21; 0.20]$ & $ [-0.14; 0.14]$ & $ [-0.14; 0.14]$ \\
X2 & $1.17^{*}$       & $2.15^{*}$       & $0.73^{*}$       & $1.18^{*}$       & $2.42^{*}$        & $0.51^{*}$        & $1.10^{*}$       & $2.47^{*}$       & $0.61^{*}$       \\
   & $ [ 1.06; 1.29]$ & $ [ 2.10; 2.20]$ & $ [ 0.64; 0.82]$ & $ [ 1.03; 1.32]$ & $ [ 2.36;  2.48]$ & $ [ 0.40;  0.63]$ & $ [ 0.89; 1.31]$ & $ [ 2.39; 2.55]$ & $ [ 0.44; 0.77]$ \\
X3 & $0.34^{*}$       & $0.56^{*}$       & $0.23^{*}$       & $0.34^{*}$       & $0.61^{*}$        & $0.11^{*}$        & $0.32^{*}$       & $0.61^{*}$       & $0.09^{*}$       \\
   & $ [ 0.29; 0.40]$ & $ [ 0.52; 0.59]$ & $ [ 0.19; 0.27]$ & $ [ 0.27; 0.41]$ & $ [ 0.57;  0.65]$ & $ [ 0.06;  0.16]$ & $ [ 0.22; 0.41]$ & $ [ 0.56; 0.67]$ & $ [ 0.02; 0.16]$ \\
X4 & $-0.00$          & $-0.00$          & $0.03$           & $-0.00$          & $-0.00$           & $-0.02$           & $-0.00$          & $-0.00$          & $-0.02$          \\
   & $ [-0.06; 0.06]$ & $ [-0.04; 0.04]$ & $ [-0.01; 0.07]$ & $ [-0.07; 0.07]$ & $ [-0.05;  0.05]$ & $ [-0.07;  0.03]$ & $ [-0.11; 0.10]$ & $ [-0.07; 0.07]$ & $ [-0.09; 0.05]$ \\
X5 & $0.16^{*}$       & $0.21^{*}$       & $0.12^{*}$       & $-0.10$          & $-0.16^{*}$       & $-0.33^{*}$       & $-0.00$          & $0.01$           & $0.00$           \\
   & $ [ 0.01; 0.30]$ & $ [ 0.11; 0.30]$ & $ [ 0.04; 0.21]$ & $ [-0.26; 0.06]$ & $ [-0.27; -0.05]$ & $ [-0.43; -0.22]$ & $ [-0.21; 0.21]$ & $ [-0.13; 0.15]$ & $ [-0.14; 0.14]$ \\
\hline
\end{tabular}
%\caption{Statistical models}
\label{table:clans}
\end{adjustbox}
\end{center}
{\footnotesize \justifying \singlespacing{This table presents CLAN differences in means for the most and least affected groups, $\hat{\delta}_K - \hat{\delta}_1$. 95\% confidence intervals are shown in brackets, with a star indicating the interval does not cover zero. } \par}
%\label{table:clans}
\end{table}

\section{Case Study}

There is a large degree of heterogeneity in Family Health Program (FHP, ESF) coverage level across time and geographic areas in Brazil, which is an opportune setting to apply a DiD with staggered timing. Other opportune characteristics are: 

\textbf{Treatment Heterogeneity:} Prior work has found heterogeneity in treatment effects of the Program for a variety of sub-populations \parencite{rocha2010evaluating,bhalotra2016does,castro2019brazil,hone2020primary}, therefore we also have an ideal candidate for application of ML methods in this setting. Hone et al (2020), for example, find that FHP utilization led to larger reductions in adverse health outcomes among more deprived racial and socioeconomic groups, and among those with lower education levels \parencite{hone2020primary}. Most other evidence from low and middle income countries does not disaggregate the effects of primary healthcare by socioeconomic groups. In earlier work, Hone et al (2017) also show that ESF expansion reduced mortality more in the black/pardo population compared to the white population \parencite{hone2017association}. Our paper potentially addresses gaps in the literature by using ML methods that can capture such treatment effect heterogeneity in a disciplined manner, for a large variety of subgroups. 

\textbf{Unobserved Confounding:} may exist through selection into the FHP. Defining treatment as the introduction of the program in a municipality (rather than percentage of the population enrolled in the program) and studying outcomes at the municipality level may help minimize related selection bias. There is a potential issue in that FHP coverage lagged behind in areas with urban poor communities. At the same time, country-wide expansion initially focused on poorer-than-average municipalities \parencite{macinko2015brazil}. For example, although FHP achieved 50\% coverage of the Brazilian population by the mid-2000s, this rate was 8 times lower in the city of Rio. This indicates potential endogeneity between introduction of treatment and the health outcomes, especially since health behaviors are related to socioeconomic status in Brazil \parencite{de2016social}. Our MLDID addresses this issue by relaxing the assumption of independence between treatment and time.  

\textbf{Program Implementation and Target Conditions:} Federal guidelines specify health conditions that ESF teams must cover and monitor: these include hypertension, diabetes, tuberculosis, and women and children's health. Since there is a large degree of variation in local implementation of the program, focusing on health outcomes related to these conditions is likely most useful in our setting. There are multiple papers on maternal health and infant mortality, and examining these outcomes will allow us to compare our findings to prior work.

\subsection{Background}

The Family Health Program is the main primary healthcare program within Brazil's national health system, the \textit{Sistema Unico de Saude}. The program is managed at the municipality level. In 1998 roughly 4\% of the population was covered by the program, and by 2014 62\% were covered \parencite{macinko2015brazil}. Increases in FHP coverage are associated with decreases in infant mortality rates \parencite{aquino2009impact,rocha2010evaluating,bhalotra2016does}. FHP users have been found to have lower likelihood of death across all causes except HIV/AIDS and maternal causes of death \parencite{hone2020primary}. Hospital admissions for primary healthcare-sensitive conditions are reduced in areas with greater FHP coverage \parencite{dourado2011trends}. 

There are other social programs in Brazil that we must consider in conjunction with the FHP. Variation in poverty and income levels across geographic areas may be mitigated by the receipt of conditional cash transfers through the Brazilian \textit{Bolsa Familia} program. Increased Bolsa Familia coverage has been shown to decrease under-5 mortality rates after adjusting for FHP coverage rates \parencite{rasella2013effect}. Other studies have examined the combined impacts of FHP coverage and the Bolsa Familia program. Therefore we ideally need to include Bolsa Familia coverage as a covariate in whichever specification we use. 

\subsection{Data}

We focus on the municipality level, the smallest administrative area in Brazil, from 1996-2008. There were 5,565 municipalities in Brazil as of 2000 and our main balanced sample includes 4,854 (some municipalities were newly created or shifted borders during the sample period, so we do not include these). Data on ESF coverage, as well as mortality and other healthcare statistics, are publicly available and taken from the Brazilian Ministry of Health (Datasus). Information on the number of ESF teams in a municipality and percent coverage per capita are not available until after 1998, and since 30\% of municipalities already had ESF coverage by this time, we do not include these covariates in the main analysis (although we plan to include them in a secondary analysis of the period from 2002-20012). The treatment is defined as the year in which a municipality first had ESF coverage.\footnote{There are some municipalities in which coverage began, paused, and then was re-implemented. We create an indicator for whether ESF coverage stopped and re-started in a municipality and control for this in the main analyses.} This information is available starting in 1996. Less than 5\% of municipalities had ESF coverage begin between 1994-1996, and we classify these municipalities as beginning coverage in 1996 (consistent with Rocha \& Soares 2010, Bhalotra et al. 2019). We also include information about \textit{Bolsa Familia} transfers. This data is available through the Ministry of Social Development. 

The Brazilian Institute of Geography and Statistics (IBGE) releases decennial census information for socioeconomic variables, as well as annual estimates of population by age and demographic group. Where sensible, Census variables have been interpolated to create annual socioeconomic indicators by municipality.\footnote{The estimation of a nuisance function for the probability of being in time period ($T=t$) requires time-varying covariates to be present in the dataset. Although this requirement is satisfied by other variables, interpolating the census data allows for the possibility that the nuisance functions are sensitive to the inclusion of time-constant $X_i$. The procedure will be repeated using only the raw decennial census variables for robustness.} These decennial IBGE data include variables related to socioeconomic development, income, and the health system, which have been shown to impact mortality rates \parencite{hone2017association}. For the main sample period (1995-2012), we include the following IBGE Census information: Gini coefficients, IDH, urbanization rates, houehold income per capita (in R\$), percentage in poverty (below R\$140) and extreme poverty (below R\$70), percentage with running water, and average literacy rates. Census data also provide information on age, racial and ethnic characteristics of municipal populations. Annual population projections for infants were stopped after 2012, therefore we end our main sample in this year. Health system variables, taken from the National System of Information on Ambulatory Care (Datasus/SIA) and the National System of Information on Hospital Care (Datasus/SIH) include immunization coverage rates, hospitalizations per capita, and outpatient procedures per capita. Healthcare expenditure per capita is taken from FINBRA and SIOPS. Mortality data for infants and adults is taken from the Brazilian National System of Mortality Records (SIM). SIM also provides detailed information about cause of mortality. Summary statistics for the baseline year 1996, as well as detailed information about data sources, are shown in Table 1. 

\textbf{Data Cleaning and Preprocessing.} In early years of the dataset, there are some obvious measurement errors with some outcome variables, such as infant mortality rate. For example, the maximum rate is reported as 15,0000 deaths per 1,000 individuals, which is a mathematically impossible amount and likely a result of administrative error in those early years.\footnote{Our secondary sample will begin in later years and avoid this problem.} To address this, we truncate any infant mortality rate above 500 to equal 500 deaths per 1000, which does not impact the distribution of this outcome and is consistent with previous work. 

%\def\sym#1{\ifmmode^{#1}\else\(^{#1}\)\fi}
%\begin{landscape}
\begin{table}[htbp]
%\begin{table}[htbp] 
    \centering
    \caption{Summary Statistics: Main Sample (1995-2012)} 
    \begin{adjustbox}{width=.9\textwidth}
    \begin{threeparttable}
    \begin{tabular}{lccccc} 
    \toprule
            %&\multicolumn{5}{c}{(1)}                                         \\
            \multicolumn{5}{c}{\textbf{Municipality Socioeconomic Characteristics} (Year = 1996)\tnote{a}} \\
%& {\textbf{Municipality Socioeconomic Characteristics}} (Year = 1996)  & & & & \\
            &        mean&          sd&         min&         max&       N  \\
    \midrule
Completed secondary education (\%)&        0.14&        0.08&        0.00&        0.54&        4972\\
Gini                &        0.54&        0.06&        0.33&        0.81&        4972\\
Gini, 2000 Census   &        0.55&        0.07&        0.30&        0.82&        4964\\
IDH                 &        0.47&        0.10&        0.20&        0.77&        4972\\
IDH, 2000 Census    &        0.53&        0.10&        0.22&        0.82&        4964\\
Urbanization rate   &        0.58&        0.22&        0.00&        1.10&        4972\\
Urbanization rate, 2000 Census&        0.62&        0.22&        0.02&        1.00&        4964\\
Income per capita, 2000 Census&      400.95&      219.56&       70.92&     2324.69&        4964\\
Extreme poverty (below R\$70)&        0.11&        0.10&        0.00&        0.86&        4972\\
Extreme poverty (below R\$70), 2000 Census&        0.18&        0.14&        0.00&        0.75&        4964\\
Poverty (below R\$140)&        0.22&        0.14&        0.00&        1.00&        4972\\
Poverty (below R\$140), 2000 Census&        0.34&        0.20&        0.01&        0.89&        4964\\
Running Water       &        0.30&        0.26&        0.00&        0.97&        4972\\
Running water, 2000 Census&        0.24&        0.29&        0.00&        0.98&        4964\\
Literacy rate       &        0.73&        0.14&        0.22&        0.98&        4972\\
Literacy rate, 2000 Census &        0.78&        0.12&        0.34&        0.98&        4964\\
White               &        0.53&        0.26&        0.02&        1.00&        4972\\
White, 2000 Census  &        0.54&        0.25&        0.01&        1.00&        4964\\
Black               &        0.05&        0.04&        0.00&        0.45&        4972\\
Black, 2000 Census  &        0.06&        0.05&        0.00&        0.42&        4964\\
Pardo               &        0.41&        0.25&        0.00&        0.96&        4972\\
Pardo, 2000 Census  &        0.40&        0.24&        0.00&        0.99&        4964\\
Indigenous          &        0.00&        0.03&        0.00&        0.78&        4972\\
Indigenous, 2000 Census&        0.01&        0.03&        0.00&        0.82&        4964\\
    \midrule \midrule 
            %&\multicolumn{5}{l}{(2)}                                         \\
            \multicolumn{5}{c}{\textbf{Health Infrastructure and Expenditure}\tnote{b}}                                            \\
            &        mean&          sd&         min&         max&       count\\
    \midrule
Immunization coverage (\%)\tnote[c]&       42.06&       25.21&        0.00&      202.88&        4972\\
Hospital present    &        0.67&        0.47&        0.00&        1.00&        4965\\
Hospitalizations per capita&        0.08&        0.05&        0.00&        0.81&        3654\\
Outpatient procedures per capita&        7.28&        5.07&        0.00&       54.72&        4972\\
Healthcare expenditure per capita&       49.76&       44.49&        0.00&      810.38&        4758\\
    \midrule
                
    \bottomrule
    \end{tabular}
    \begin{tablenotes}
        \item[a] Demographic and Socioeconomic variables taken from Datasus and the Decennial Census. All variables were linearly interpolated to create annual indicators. "2000 Census" denotes raw value from the 2000 year Census data, included for comparison.
        \item[b] Health indicators taken from Datasus/SIA. Health financing expenditure is a combined measure of Finbra and SIOPS. Finbra is used for 1996 and 1997. SIOPS for 2000 and 2001. For all years past 2001, the average of the two estimates is used. Expenditures for 1995 are linearly interpolated from 1996 estimates. 
        \item[c] Immunization coverage from the National Immunization Program Information System and Datasus. It includes Hepatitis A and B, Rotavirus, Polio, DTP, Adult DTP, Measles and Influenza. Rates over 100\% should indicate boosters or secondary/seasonal coverage. 
    \end{tablenotes}
    \end{threeparttable}
    \end{adjustbox}
\end{table}
%\end{landscape}
%\end{table}

\subsection{Results: Group-Time ATTs}

Our first sample period covers the years 1996-2008. For the infant mortality outcome, there are 4,564 municipalities in the sample. We control for presence of a hospital, percent urban, percent with secondary education,\footnote{As the method requires some degree of variation in covariates across municipalities, it is not possible to include a control for the bolsa familia program as in \textcite{bhalotra2016does}. As enrollment in Bolsa Familia is correlated with poverty rates, we include the urban and demographic covariates as proxies for future BF coverage. We also address this concern in the secondary sample (see next section).} and a dummy for the Southeast region (the richest region, containing both São Paulo and Rio de Janeiro). 

\begin{figure}[H]
\caption{Comparison of Methods: Case Study}
    \centering
    \includegraphics{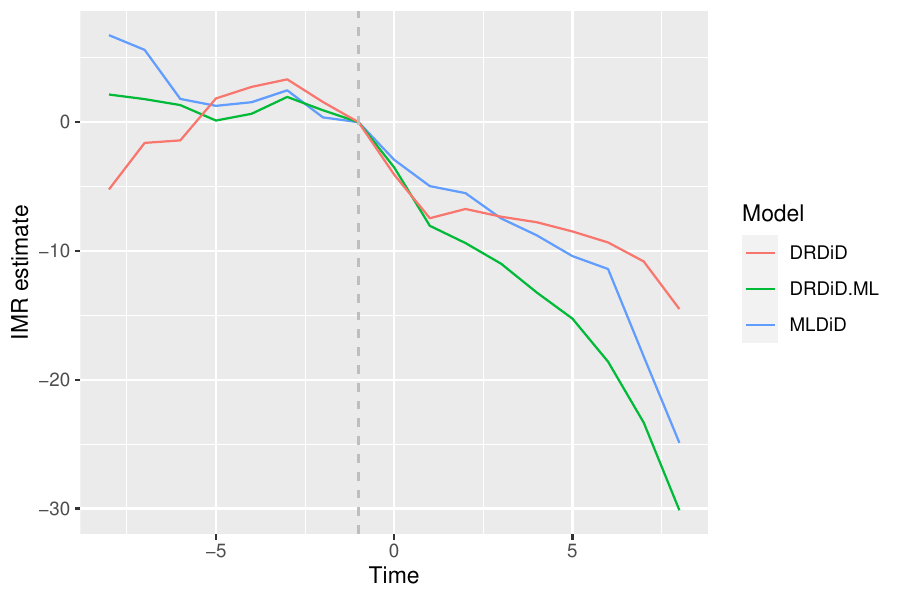}
    %\caption{Caption}
    \label{fig:methodscompare}
\end{figure}

Figure \ref{fig:methodscompare} depicts the aggregated dynamic ATTs for our method, MLDID, and two versions of the DRDID of \textcite{callaway2021difference}, one implemented within our own loop structure (see Algorithm 1) and the other taken from the {\tt R did} package. Our MLDID estimates are close to both versions of DRDID, especially in the times close to treatment. Note that we use time $t = -1$ as the reference time period, indicated by the dotted line. 

Next, we compare our dynamic MLDID ATT estimates to the DRDID dynamic estimates in Figure \ref{dynatt_psf} (point estimates are reported in Appendix Table \ref{table:dynamiccoefficients}). Both versions show a negative and significant effect of the implementation of PSF in a municipality on infant mortality. The MLDID estimates are only significantly below zero beginning a few years after treatment, while the DRDID estimates are immediately negatively significant.  

\begin{figure}[H]
    \centering
    \begin{subfigure}{0.48\textwidth}
        \centering
        \includegraphics[width=\linewidth]{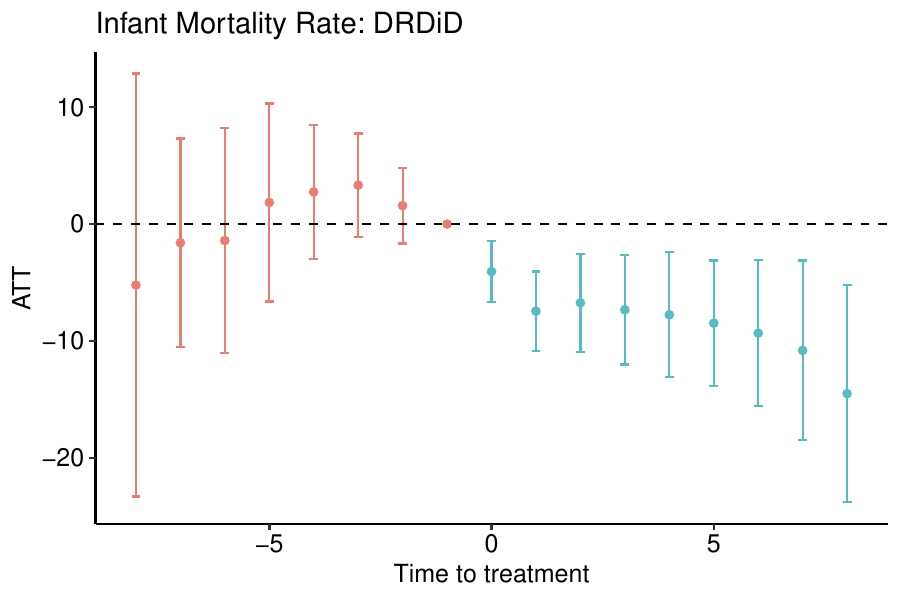}
        \caption{DRDID (Callaway Sant'Anna)}
    \end{subfigure}
    \quad
    \begin{subfigure}{0.48\textwidth}
        \centering
        \includegraphics[width=\linewidth]{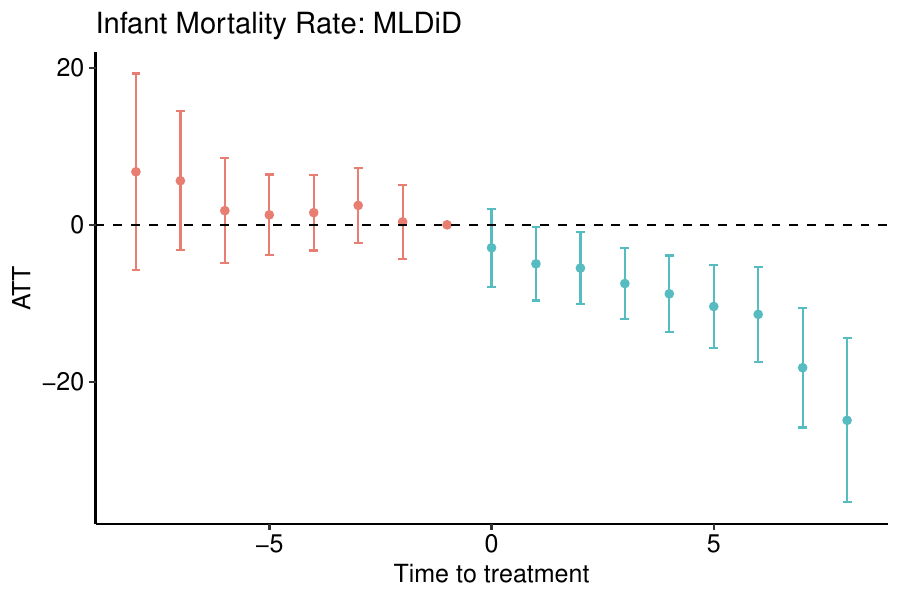}
        \caption{MLDID (Hatamyar et al)}
    \end{subfigure}
    \caption{Effect of PSF on Infant Mortality}
    \label{dynatt_psf}
    {\footnotesize \justifying \singlespacing{This figure depicts the dynamic effect (aggregated ATTs) of the PSF on infant mortality. Panel A depicts our MLDID method and Panel B the DRDID of \textcite{callaway2021difference}. Point estimates and standard errors are reported in Appendix Table \ref{table:dynamiccoefficients}. } \par}
\end{figure}

\subsection{Treatment Effect Heterogeneity}

\begin{figure}[H]
    \caption{Dynamic CATEs: Effect of FHP on Infant Mortality}
    \centering
    \includegraphics{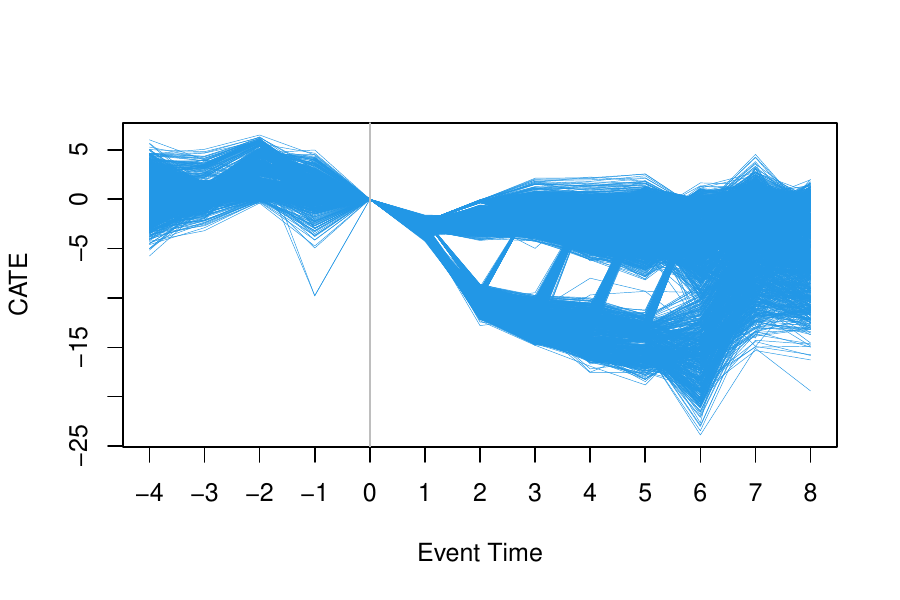}
    \label{fig:dynamiccates}
    {\footnotesize \singlespacing{This figure depicts the individual dynamic CATEs, where each line represents the estimated CATE for a single municipality.} \par}
\end{figure}

To examine treatment effect heterogeneity, we first plot the dynamic CATEs in Figure \ref{fig:dynamiccates}. We can see a large degree of heterogeneity in the CATE estimates, especially as time since treatment increases. 

\begin{figure}[H]
    \centering
    \begin{subfigure}{0.48\textwidth}
        \centering
        \includegraphics[width=\linewidth]{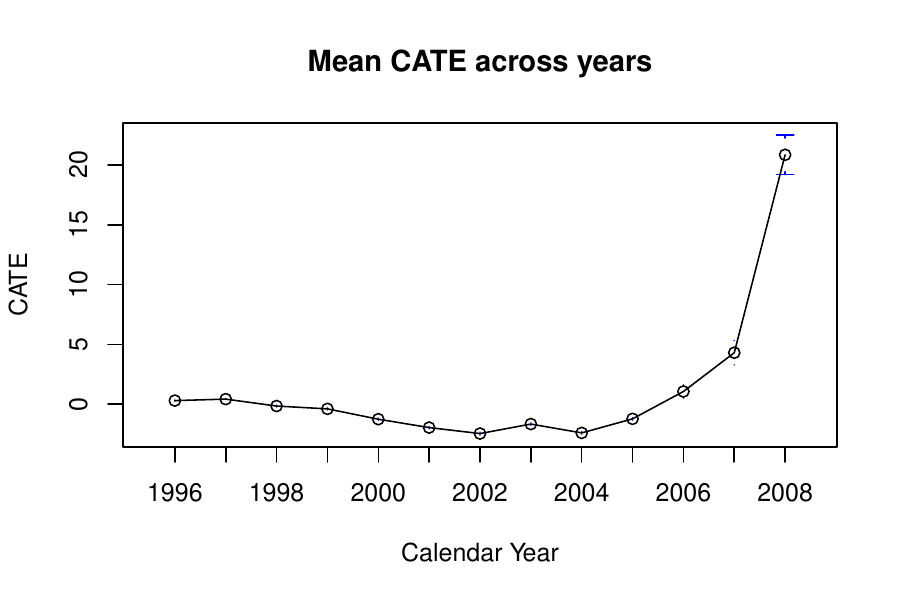}
        %\caption{DRDID (Callaway Sant'Anna)}
    \end{subfigure}
    \quad
    \begin{subfigure}{0.48\textwidth}
        \centering
        \includegraphics[width=\linewidth]{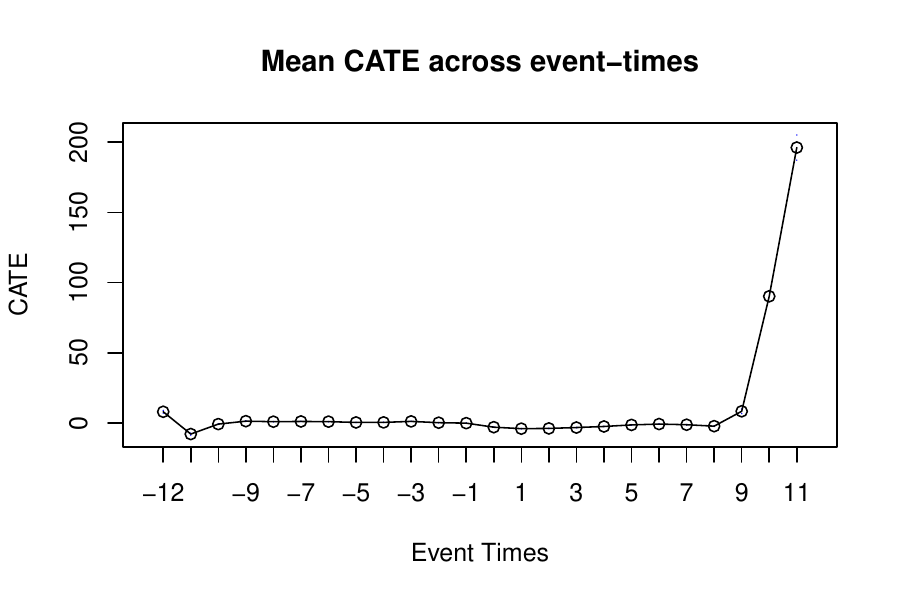}
        %\caption{MLDID (Hatamyar et al)}
    \end{subfigure}
    \caption{Stability of CATT estimates}
    \label{fig:stability}
\end{figure}

We also examine the stability of the estimator with respect to event time and calendar time in Figure \ref{fig:stability}. It is clear that the estimator is unstable in later years and in later event-times. The later event times (Panel B) are only possible to estimate for those municipalities treated at the beginning of the sample period, and the control group for those later times will consist of a smaller and smaller number of observations (as most municipalities will have moved out of the control group by the end of the sample). This issue is also inherent in other parametric and nonparametric staggered DID models. We also see that the mean CATE for later years in the sample is highly unstable, for similar reasons (most municipalities would have been treated by the end of the sample). 

\subsection{Best Linear Predictors of Treatment Effect Heterogeneity}

\begin{figure}[H]
    \centering\includegraphics{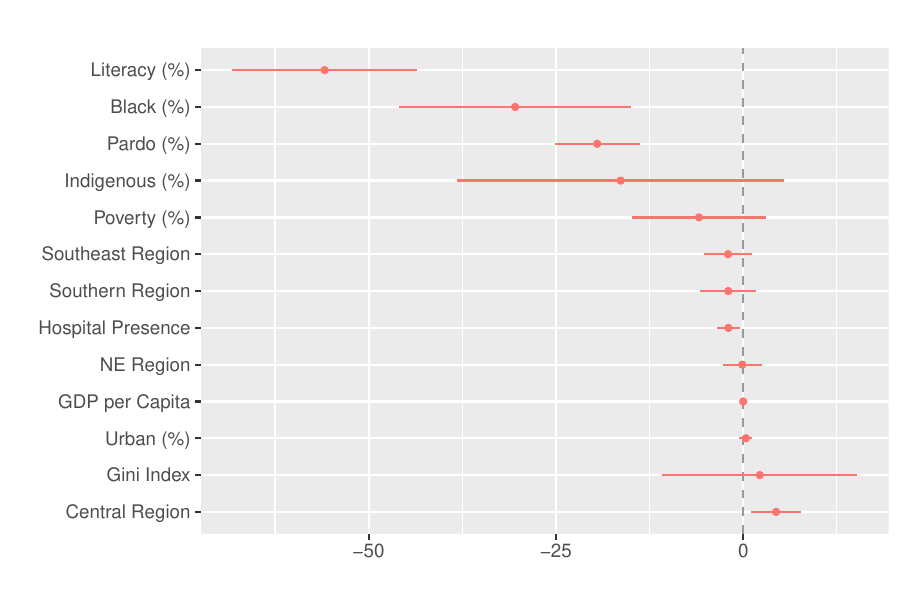}
    \caption{Best Linear Predictor of Heterogeneity}
    \label{fig:BLP_pooled}
    {\footnotesize \singlespacing{This figure depicts a single time period since treatment, $e = 1$, of the BLP regression coefficients for our case study. The rows indicate the individual covariate effect on the estimated DR scores, at this time period.} \par}
\end{figure}

We first examine the BLP of our treatment effect heterogeneity by regressing the $\hat{\tau}(x)$, as well as the ``scores" $\hat{\gamma}(X_i, G_i, T_i)(Y_i - \hat{y}(X_i, G_i, T_i))$, on the covariates in the data. Each estimate (CATE or score) is matched to the year of the data corresponding to the event time (time since treatment). Table \ref{table:BLPaverage} depicts OLS estimates for this overall BLP analysis, using a variety of specifications, and Figure \ref{fig:BLP_pooled} shows the coefficients for the first year of the sample only. Coefficients indicate how much treatment effect heterogeneity changes over time, as each regressor (covariate) increases. Our simulation results indicate that the score-based BLP is closer to the true impact of covariates on heterogeneity than the CATE-based BLP, but we include both here for comparison. Over the entire sample period, we see that urban locations significantly predict negative treatment effect heterogeneity for both BLP types. Poverty has a significant negative effect on the scores in all specifications, indicating that, overall, those in poorer areas experienced a greater reduction in infant mortality due to FHP implementation. We also see the effect of supply-side variables, such as hospital presence and proportion of population covered by healthcare teams, as having a small and significant negative impact on the scores. Finally, we can see that geographic regions have some positive impact on treatment effect heterogeneity (compared to the baseline Northern region). 

These estimates do not tell the whole story, however, and our novel estimation procedure allows us to disentangle the effects of these covariates on treatment effect heterogeneity in a dynamic way. Figure \ref{fig:dynamicBLPs} presents coefficients from an OLS where the score is regressed onto a selection of covariates for subsets of the data in each time since treatment (that is, for each relative $e$). Point estimates corresponding to these individual BLPs are presented in Table \ref{table:BLPdynamicPSF}. We can see that, for many covariates, the impact on treatment effect heterogeneity only begins a few years after the implementation of the PSF. In the first year of implementation (column (0) in Table \ref{table:BLPdynamicPSF}), only the percent of individuals in poverty and the southeast region significantly predict treatment effect heterogeneity. By 8 years after treatment (column (8)), ten out of sixteen covariates show a significant relationship with treatment effect heterogeneity. The dynamic graphs in Figure \ref{fig:dynamicBLPs} show that the overall dynamic treatment effect of FHP on infant mortality is heterogeneous in a dynamic manner.

%\textbf{NEED TO ADD DESCRIPTION/SOME INTERPRETATION OF THESE RESULTS}. 
%The dynamic BLP estimates show us that those negative effects on heterogeneity for covariates such as inequality index (Panel A) are only present starting 5 years after treatment. The presence of a hospital (Panel D) actually has a negative and insignificant effect on treatment effect heterogeneity, but only for the first few years after treatment. Panel F shows that the prosperous Southeast region effect on treatment effect heterogeneity only gradually becomes a positive effect over time. Point estimates for these individual BLPs are presented in Table \ref{table:BLPdynamic}. 

\begin{table}[!htbp] \centering 
  \caption{Best Linear Predictors of Heterogeneity: Entire Sample} 
  \label{table:BLPaverage} 
  \begin{adjustbox}{max width=0.9\textwidth, max height=0.9\textheight}
\begin{tabular}{@{\extracolsep{5pt}}lcccccc} 
\\[-1.8ex]\hline 
\hline \\[-1.8ex] 
 & \multicolumn{6}{c}{\textit{Dependent variable:}} \\ 
\cline{2-7} 
\\[-1.8ex] & CATE & score & CATE & score & CATE & score \\ 
\\[-1.8ex] & (1) & (2) & (3) & (4) & (5) & (6)\\ 
\hline \\[-1.8ex] 
 Gini Index & $-$1.991 & $-$0.998 & $-$0.190 & 3.175 & $-$3.134 & $-$1.225 \\ 
  & (4.034) & (6.627) & (4.037) & (6.637) & (4.102) & (6.745) \\ 
  & & & & & & \\ 
 \% Urban & 1.092$^{***}$ & 1.687$^{***}$ & 1.126$^{***}$ & 1.928$^{***}$ & 0.934$^{***}$ & 1.641$^{***}$ \\ 
  & (0.286) & (0.469) & (0.290) & (0.476) & (0.293) & (0.482) \\ 
  & & & & & & \\ 
 Income per-capita & $-$0.000 & $-$0.00000 & 0.000 & $-$0.000 & 0.000 & $-$0.00000 \\ 
  & (0.000) & (0.00000) & (0.000) & (0.00000) & (0.000) & (0.00000) \\ 
  & & & & & & \\ 
 Poverty (140/day) & $-$6.697$^{**}$ & $-$12.261$^{***}$ & 0.979 & $-$15.013$^{***}$ & 2.768 & $-$12.339$^{***}$ \\ 
  & (2.705) & (4.444) & (2.848) & (4.683) & (2.882) & (4.739) \\ 
  & & & & & & \\ 
 \% Literate & 18.330$^{***}$ & $-$27.003$^{***}$ & 5.058 & $-$21.370$^{***}$ & 5.836 & $-$20.208$^{***}$ \\ 
  & (3.916) & (6.433) & (4.508) & (7.411) & (4.511) & (7.417) \\ 
  & & & & & & \\ 
 \% Black & 11.969$^{**}$ & $-$33.836$^{***}$ & 10.745$^{**}$ & $-$32.140$^{***}$ & 10.680$^{**}$ & $-$32.237$^{***}$ \\ 
  & (4.648) & (7.636) & (4.793) & (7.880) & (4.792) & (7.879) \\ 
  & & & & & & \\ 
 \% Pardo & $-$5.388$^{***}$ & $-$11.202$^{***}$ & $-$2.090 & $-$15.961$^{***}$ & $-$1.525 & $-$15.118$^{***}$ \\ 
  & (1.476) & (2.425) & (1.776) & (2.920) & (1.782) & (2.929) \\ 
  & & & & & & \\ 
 \% Indigenous & $-$4.466 & $-$7.319 & $-$3.284 & $-$10.391 & $-$2.731 & $-$9.565 \\ 
  & (6.654) & (10.931) & (6.776) & (11.140) & (6.776) & (11.141) \\ 
  & & & & & & \\ 
 Hospital & 0.034 & $-$1.422$^{*}$ & 0.905$^{*}$ & $-$1.652$^{**}$ & 0.694 & $-$1.967$^{**}$ \\ 
  & (0.473) & (0.777) & (0.479) & (0.787) & (0.482) & (0.792) \\ 
  & & & & & & \\ 
 \% Secondary Education & $-$73.329$^{***}$ & $-$55.783$^{***}$ & $-$71.526$^{***}$ & $-$63.190$^{***}$ & $-$65.997$^{***}$ & $-$54.927$^{***}$ \\ 
  & (4.405) & (7.237) & (4.352) & (7.155) & (4.565) & (7.505) \\ 
  & & & & & & \\ 
 Propotion ESF  & $-$0.024$^{***}$ & $-$0.028$^{***}$ &  &  & $-$0.022$^{***}$ & $-$0.033$^{***}$ \\ 
  & (0.006) & (0.009) &  &  & (0.006) & (0.009) \\ 
  & & & & & & \\ 
 NE &  &  & 0.783 & 1.757 & 1.085 & 2.210 \\ 
  &  &  & (0.823) & (1.353) & (0.826) & (1.358) \\ 
  & & & & & & \\ 
 CO &  &  & 7.370$^{***}$ & 2.354 & 7.726$^{***}$ & 2.887$^{*}$ \\ 
  &  &  & (1.036) & (1.703) & (1.039) & (1.709) \\ 
  & & & & & & \\ 
 SoE &  &  & 9.474$^{***}$ & $-$2.600 & 9.476$^{***}$ & $-$2.597 \\ 
  &  &  & (0.979) & (1.610) & (0.979) & (1.610) \\ 
  & & & & & & \\ 
 S &  &  & 7.927$^{***}$ & $-$2.311 & 8.001$^{***}$ & $-$2.200 \\ 
  &  &  & (1.155) & (1.899) & (1.155) & (1.899) \\ 
  & & & & & & \\ 
 Constant & 2.741 & 42.027$^{***}$ & 1.327 & 38.919$^{***}$ & 1.439 & 39.087$^{***}$ \\ 
  & (3.701) & (6.080) & (4.296) & (7.063) & (4.295) & (7.062) \\ 
  & & & & & & \\ 
\hline \\[-1.8ex] 
Observations & 56,680 & 56,680 & 56,680 & 56,680 & 56,680 & 56,680 \\ 
R$^{2}$ & 0.009 & 0.005 & 0.011 & 0.005 & 0.011 & 0.005 \\ 
Adjusted R$^{2}$ & 0.009 & 0.005 & 0.011 & 0.005 & 0.011 & 0.005 \\ 
\hline 
\hline \\[-1.8ex] 
\textit{Note:}  & \multicolumn{6}{r}{$^{*}$p$<$0.1; $^{**}$p$<$0.05; $^{***}$p$<$0.01} \\ 
\end{tabular} 
\end{adjustbox}
\end{table} 

\begin{figure}[H]
    \centering
    \caption{Dynamic BLP of Treatment Effect Heterogeneity: Family Health Programme and Infant Mortality.}\label{fig:dynamicBLPs}
    
    \begin{subfigure}[b]{0.3\textwidth}
        \centering
        \includegraphics[width=50mm]{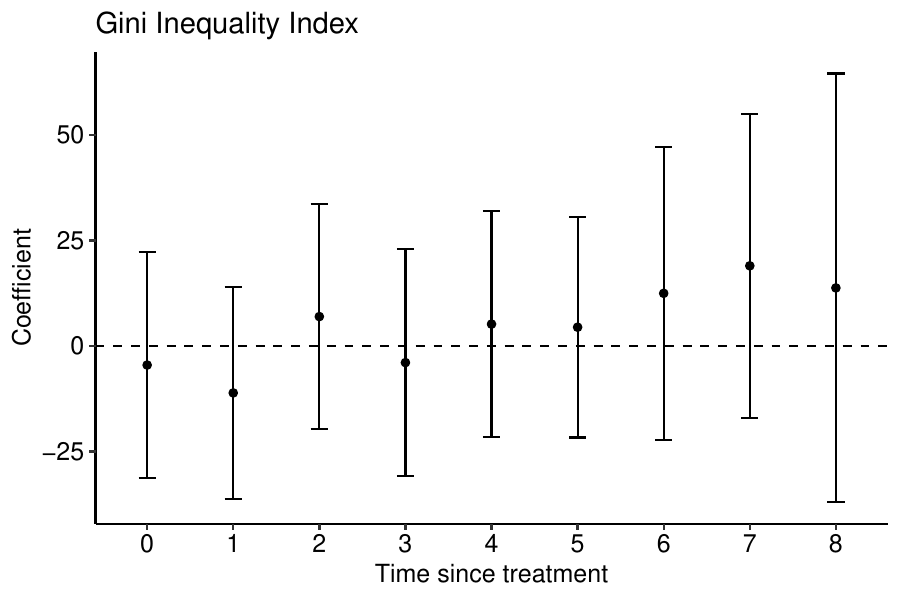}
        \caption{Gini Index}
    \end{subfigure}
    \hfill
    \begin{subfigure}[b]{0.3\textwidth}
        \centering
        \includegraphics[width=50mm]{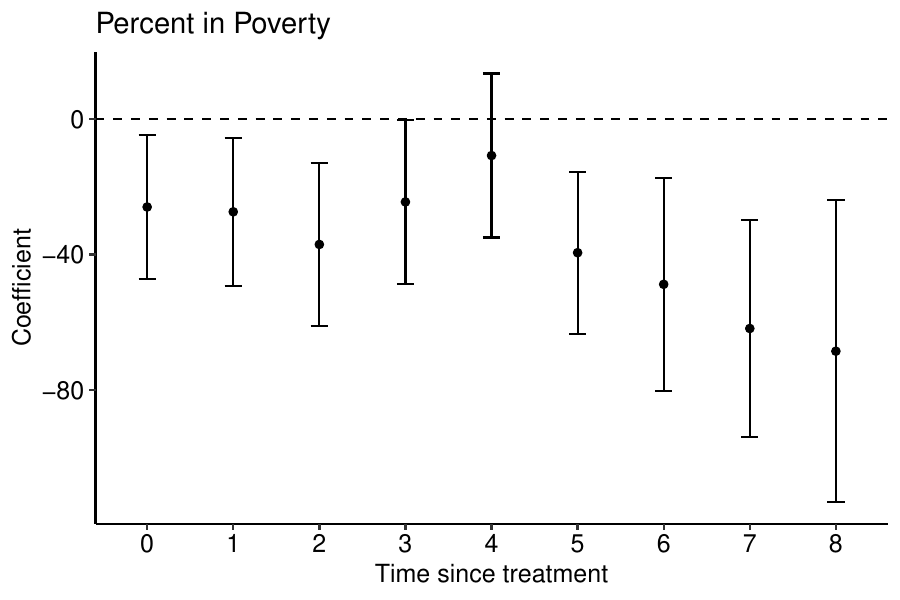}
        \caption{Percent in Poverty}
    \end{subfigure}
    \hfill
    \begin{subfigure}[b]{0.3\textwidth}
        \centering
        \includegraphics[width=50mm]{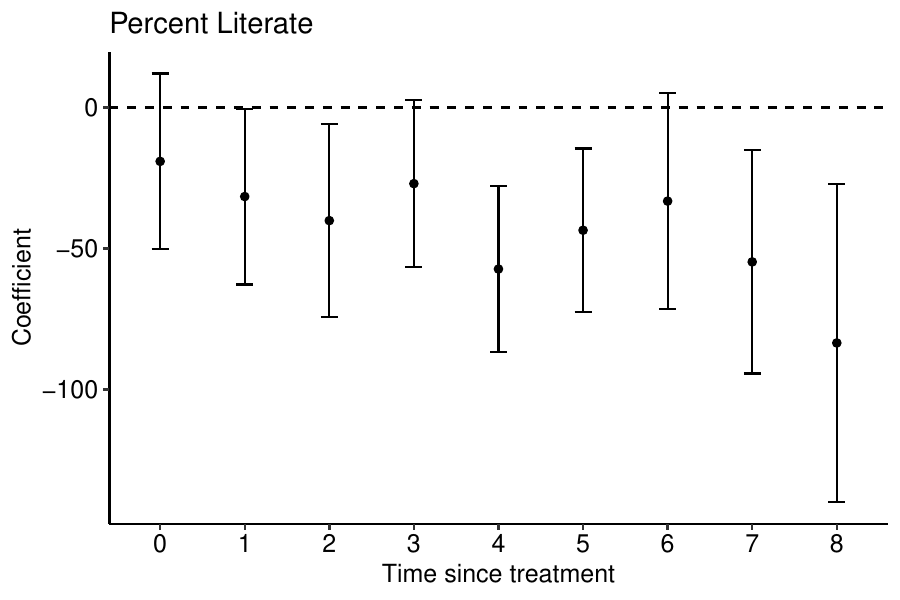}
        \caption{Percent Literate}
    \end{subfigure}

    \vspace{1em}

    \begin{subfigure}[b]{0.3\textwidth}
        \centering
        \includegraphics[width=50mm]{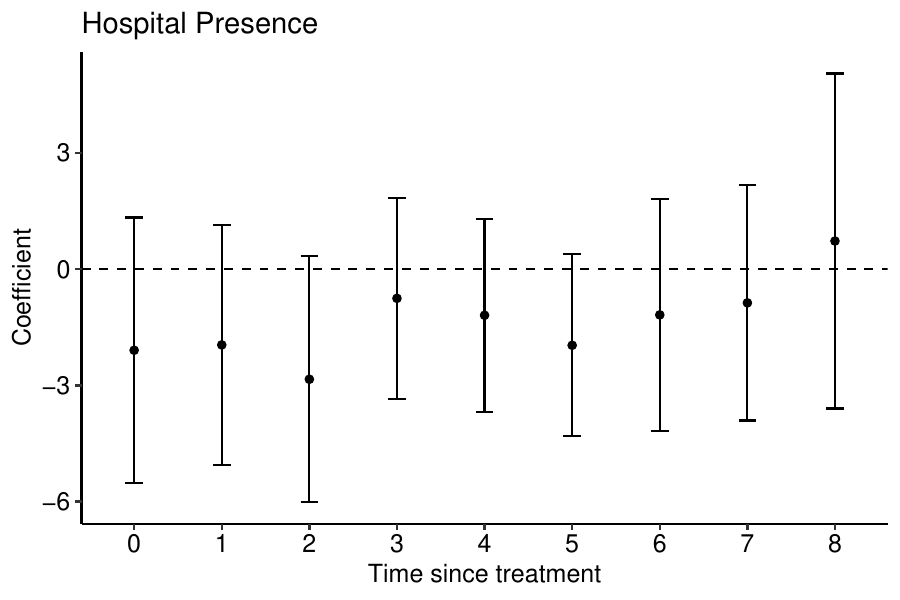}
        \caption{Hospital}
    \end{subfigure}
    \hfill
    \begin{subfigure}[b]{0.3\textwidth}
        \centering
        \includegraphics[width=50mm]{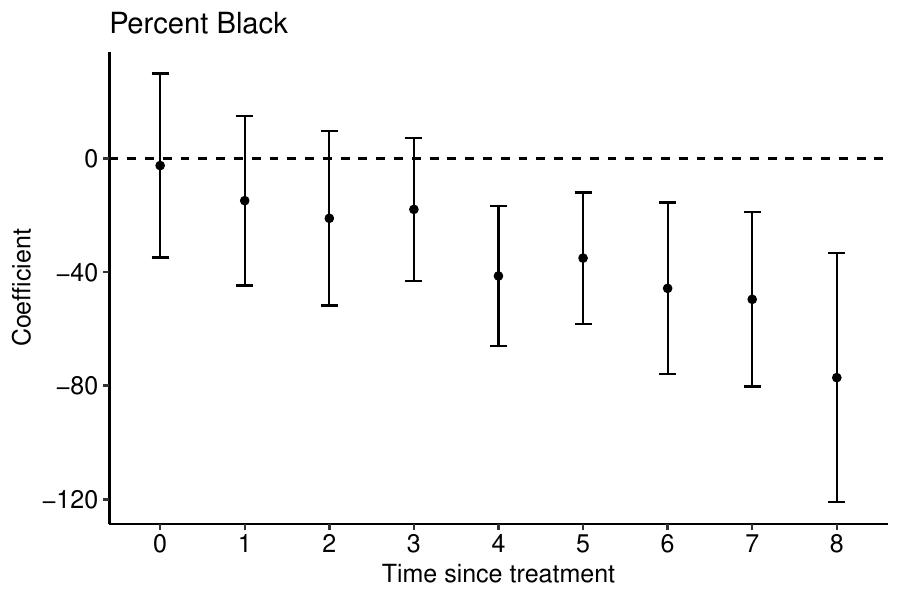}
        \caption{Percent Black}
    \end{subfigure}
    \hfill
    \begin{subfigure}[b]{0.3\textwidth}
        \centering
        \includegraphics[width=50mm]{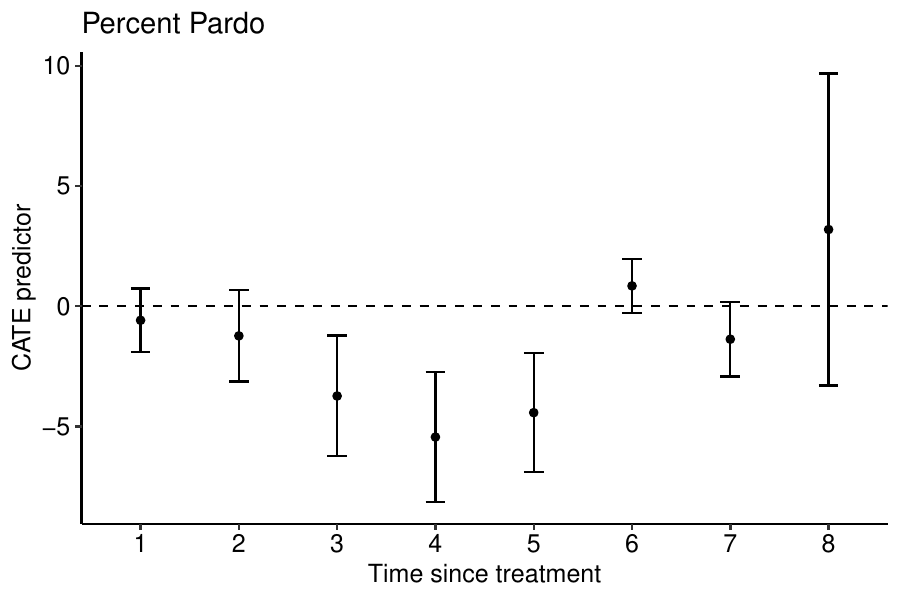}
        \caption{Percent Pardo}
    \end{subfigure}

    \vspace{1em}

    \begin{subfigure}[b]{0.3\textwidth}
        \centering
        \includegraphics[width=50mm]{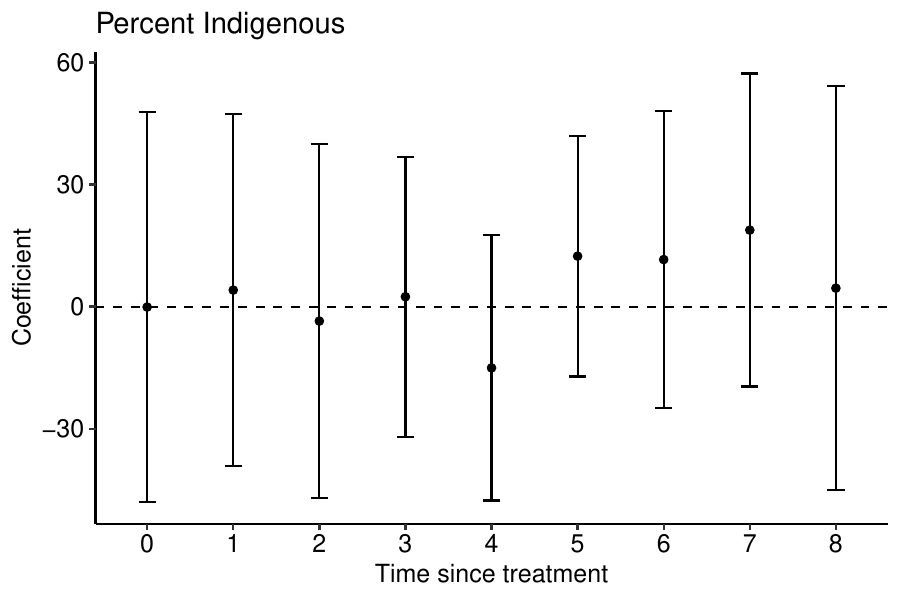}
        \caption{Percent Indigenous}
    \end{subfigure}
    \hfill
    \begin{subfigure}[b]{0.3\textwidth}
        \centering
        \includegraphics[width=50mm]{Images/dynamicBLP_hospital.pdf}
        \caption{Hospital}
    \end{subfigure}
    \hfill
    \begin{subfigure}[b]{0.3\textwidth}
        \centering
        \includegraphics[width=50mm]{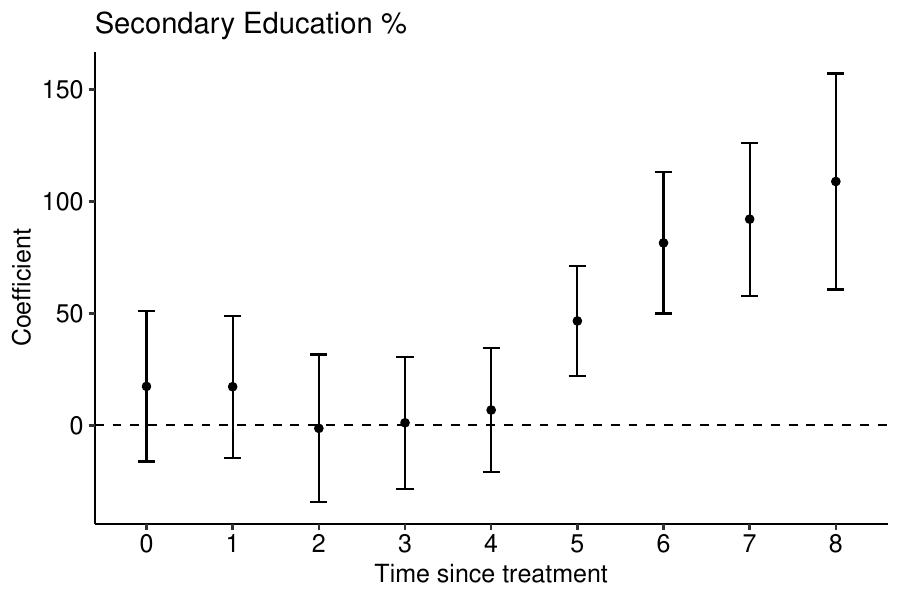}
        \caption{Secondary Education}
    \end{subfigure}

    \vspace{1em}

    \begin{subfigure}[b]{0.3\textwidth}
        \centering
        \includegraphics[width=50mm]{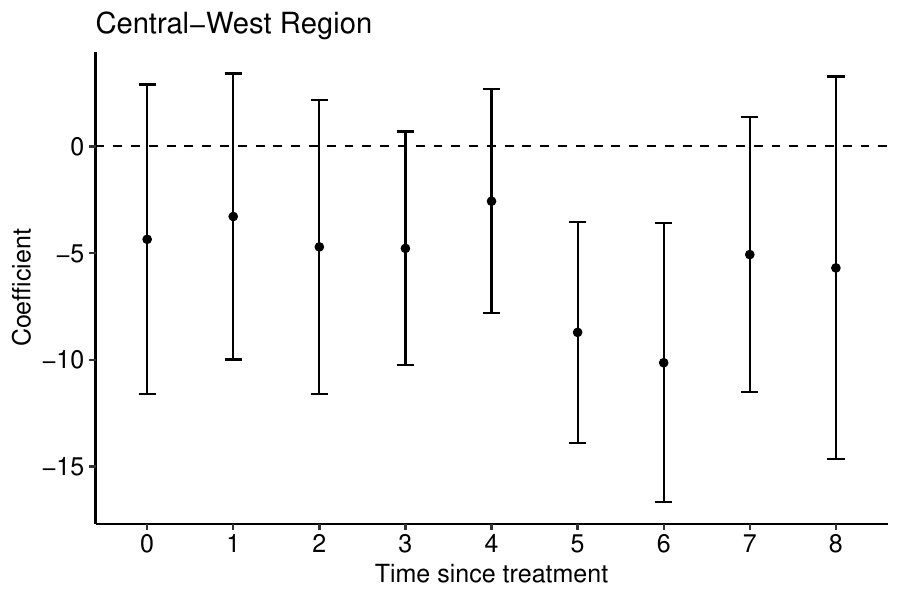}
        \caption{Central-West Region}
    \end{subfigure}
    \hfill
    \begin{subfigure}[b]{0.3\textwidth}
        \centering
        \includegraphics[width=50mm]{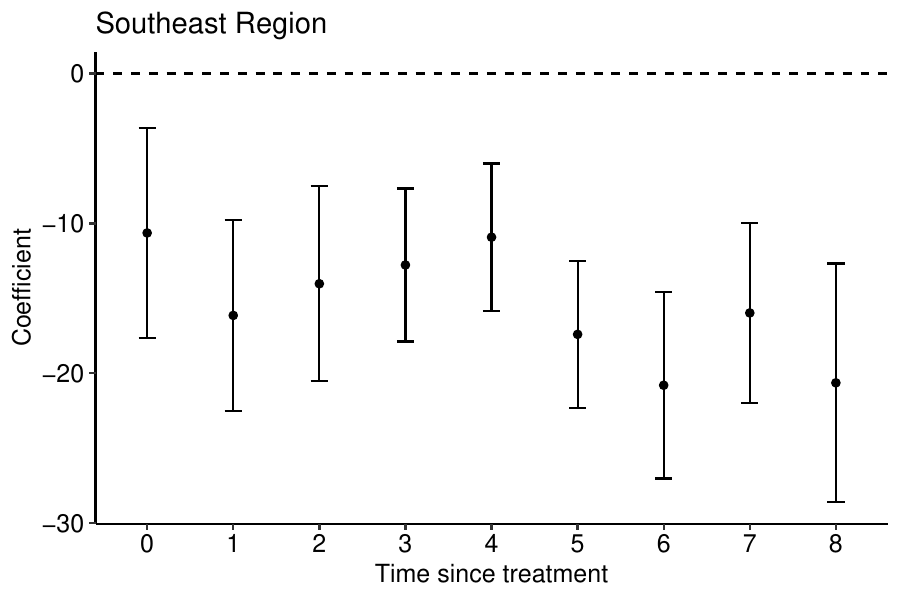}
        \caption{Southeast Region}
    \end{subfigure}
    \hfill
    \begin{subfigure}[b]{0.3\textwidth}
        \centering
        \includegraphics[width=50mm]{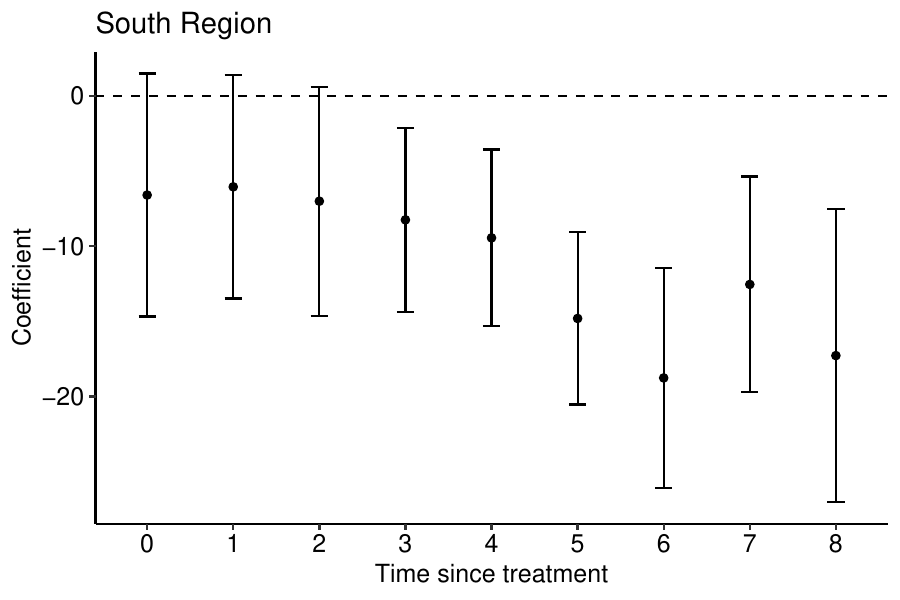}
        \caption{South Region}
    \end{subfigure}
    
    {\footnotesize \justifying \singlespacing{This figure depicts a selection of dynamic best linear predictors of heterogeneous treatment effects. This is obtained by regressing the estimated score on the data covariates, using the subsample of data for each year post-treatment. } \par}
\end{figure}

\subsection{CLANs}

By applying the Classification Analysis described earlier, we can check whether there is heterogeneity within each variable according to time since treatment. Note this analysis does not take into account other covariates and is merely a univariate way to examine heterogeneity, by comparing the difference between those most and least affected. Results in Figure \ref{fig:CLANS_PSFexample} suggest that areas with a lower literacy rate saw a greater treatment effect initially ($\hat{\delta}_K - \hat{\delta}_1 < 0$, Panel A), and those with higher poverty rate saw the greatest initial reduction in infant mortality ($\hat{\delta}_K - \hat{\delta}_1 > 0$, Panel B). These results are in line with the BLP findings, and with previous research, but provide a more granular result on the rate of change in effectiveness for the two groups. Further CLAN results are forthcoming, and differences in means for other selected covariates are in the Appendix Table \ref{table:CLANdynamicPSF}. 

\begin{figure}[H]
    \centering
    \caption{CLANs for Selected Covariates}
    \addtocounter{figure}{-1}
\begin{subfigure}{0.48\linewidth}
  \centering
  \includegraphics[width=\linewidth]{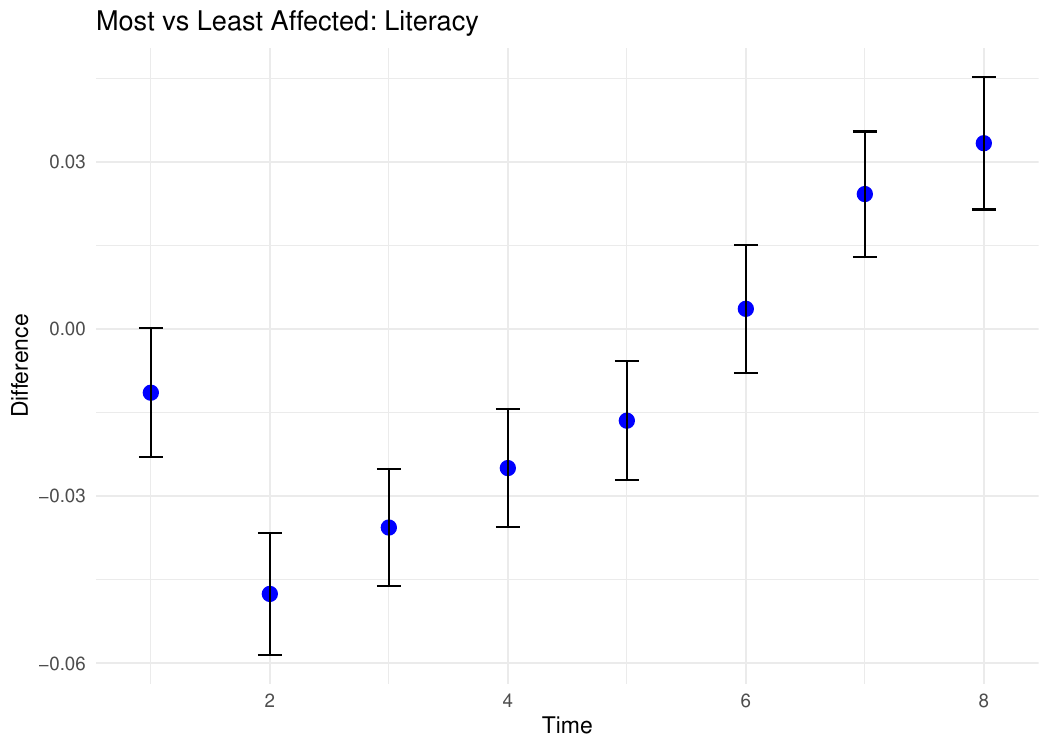}
  \caption{Literacy Rate}
  %\label{fig: Image1} 
\end{subfigure}
\quad
\begin{subfigure}{0.48\linewidth}
  \centering
  \includegraphics[width=\linewidth]{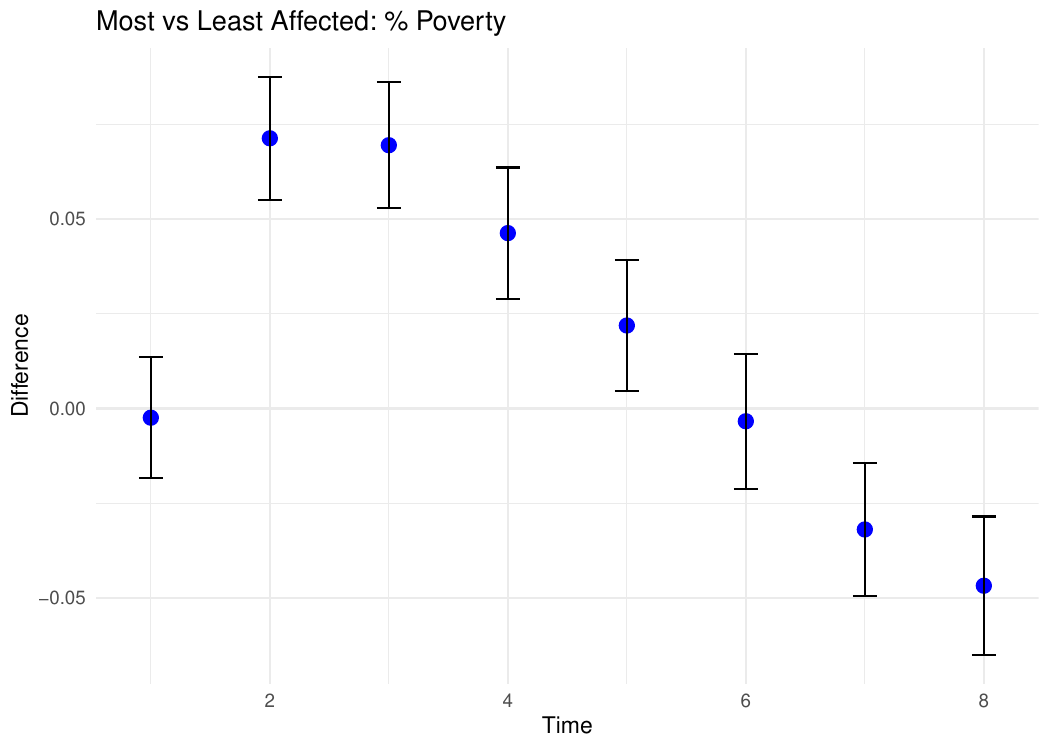}
  \caption{Poverty Rate}
  %\label{fig:NDRvaluecalc} 
\end{subfigure}
%{\footnotesize \justifying \singlespacing{This figure depicts overlap scenarios for $\psi \in (1,2)$ } \par}
\caption*{\footnotesize{This figure depicts results for a test of differences in means $\hat{\delta}_K - \hat{\delta}_1$ between the most and least affected groups, for two covariates. A positive value indicates those with a higher level of the covariate experienced a greater treatment effect, and a negative value indicates those with lower levels experienced a greater treatment effect.}}
\label{fig:CLANS_PSFexample}
\end{figure} 

\section{Conclusions}

In this paper, we implement a version of DID with staggered treatment adoption that allows us to use ML to flexibly estimate conditional average treatment effects on the treated. We then use these CATTs to examine drivers of treatment effect heterogeneity over time. We apply the method to simulated data and find that overall dynamic ATTs are consistent with other methods and the oracle ATTs, and we also find that our estimated CATTs and robust scores are able to accurately pinpoint the drivers of treatment effect heterogeneity. That is, by simulating data in which we know which covariates do truthfully drive treatment effect heterogeneity, we are able to show that our estimates pick up on those covariates using best linear predictor (BLP) and classification analysis (CLAN). Although the RMSE for individual dynamic CATT estimates is lower than the RMSE for estimated scores, we find that the CATT estimates may by more likely erroneously picking up on confounding covariates in BLP analysis -- therefore for any heterogeneity analysis we recommend using scores. We assessed a case study of the introduction of the Family Health Programme in Brazil and its impact on infant mortality rates. We were able to flexibly examine drivers of treatment effect heterogeneity, pointing to the longer-term effect of inequality, poverty, and ethnicity on reduction of infant mortality after FHP implementation. 

\clearpage

\printbibliography[heading=none]

@article{hone2020primary,
  title={Primary healthcare expansion and mortality in Brazil’s urban poor: A cohort analysis of 1.2 million adults},
  author={Hone, Thomas and Saraceni, Valeria and Medina Coeli, Claudia and Trajman, Anete and Rasella, Davide and Millett, Christopher and Durovni, Betina},
  journal={PLoS medicine},
  volume={17},
  number={10},
  pages={e1003357},
  year={2020},
  publisher={Public Library of Science San Francisco, CA USA}
}

@article{rasella2013effect,
  title={Effect of a conditional cash transfer programme on childhood mortality: a nationwide analysis of Brazilian municipalities},
  author={Rasella, Davide and Aquino, Rosana and Santos, Carlos AT and Paes-Sousa, R{\^o}mulo and Barreto, Mauricio L},
  journal={The lancet},
  volume={382},
  number={9886},
  pages={57--64},
  year={2013},
  publisher={Elsevier}
}

@article{d2022nonparametric,
  title={Nonparametric difference-in-differences in repeated cross-sections with continuous treatments},
  author={D’Haultf{\oe}uille, Xavier and Hoderlein, Stefan and Sasaki, Yuya},
  journal={Journal of Econometrics},
  year={2022},
  publisher={Elsevier}
}

@article{gavrilova2023dynamic,
  title={Dynamic Causal Forests, with an Application to Payroll Tax Incidence in Norway},
  author={Gavrilova, Evelina and Lang{\o}rgen, Audun and Zoutman, Floris},
  year={2023},
  publisher={CESifo Working Paper}
}

@techreport{chernozhukov2018generic,
  title={Generic machine learning inference on heterogeneous treatment effects in randomized experiments, with an application to immunization in India},
  author={Chernozhukov, Victor and Demirer, Mert and Duflo, Esther and Fernandez-Val, Ivan},
  year={2018},
  institution={National Bureau of Economic Research}
}

@article{baker2022much,
  title={How much should we trust staggered difference-in-differences estimates?},
  author={Baker, Andrew C and Larcker, David F and Wang, Charles CY},
  journal={Journal of Financial Economics},
  volume={144},
  number={2},
  pages={370--395},
  year={2022},
  publisher={Elsevier}
}

@article{wager2018estimation,
  title={Estimation and inference of heterogeneous treatment effects using random forests},
  author={Wager, Stefan and Athey, Susan},
  journal={Journal of the American Statistical Association},
  volume={113},
  number={523},
  pages={1228--1242},
  year={2018},
  publisher={Taylor \& Francis}
}

@article{bertrand2004much,
  title={How much should we trust differences-in-differences estimates?},
  author={Bertrand, Marianne and Duflo, Esther and Mullainathan, Sendhil},
  journal={The Quarterly journal of economics},
  volume={119},
  number={1},
  pages={249--275},
  year={2004},
  publisher={MIT Press}
}

@book{wooldridge2010econometric,
  title={Econometric analysis of cross section and panel data},
  author={Wooldridge, Jeffrey M},
  year={2010},
  publisher={MIT press}
}

@article{robinson1988root,
  title={Root-N-consistent semiparametric regression},
  author={Robinson, Peter M},
  journal={Econometrica: Journal of the Econometric Society},
  pages={931--954},
  year={1988},
  publisher={JSTOR}
}

@article{nie2021quasi,
  title={Quasi-oracle estimation of heterogeneous treatment effects},
  author={Nie, Xinkun and Wager, Stefan},
  journal={Biometrika},
  volume={108},
  number={2},
  pages={299--319},
  year={2021},
  publisher={Oxford University Press}
}

@article{hirshberg2021augmented,
  title={Augmented minimax linear estimation},
  author={Hirshberg, David A and Wager, Stefan},
  journal={The Annals of Statistics},
  volume={49},
  number={6},
  pages={3206--3227},
  year={2021},
  publisher={Institute of Mathematical Statistics}
}

@article{sant2020doubly,
  title={Doubly robust difference-in-differences estimators},
  author={Sant’Anna, Pedro HC and Zhao, Jun},
  journal={Journal of Econometrics},
  volume={219},
  number={1},
  pages={101--122},
  year={2020},
  publisher={Elsevier}
}

@article{baker2021much,
  title={How Much Should We Trust Staggered Difference-In-Differences Estimates?},
  author={Baker, Andrew and Larcker, David F and Wang, Charles CY},
  journal={Available at SSRN 3794018},
  year={2021}
}

@article{callaway2021difference,
  title={Difference-in-differences with multiple time periods},
  author={Callaway, Brantly and Sant’Anna, Pedro HC},
  journal={Journal of Econometrics},
  volume={225},
  number={2},
  pages={200--230},
  year={2021},
  publisher={Elsevier}
}

@article{de2020two,
  title={Two-way fixed effects estimators with heterogeneous treatment effects},
  author={De Chaisemartin, Cl{\'e}ment and d'Haultfoeuille, Xavier},
  journal={American Economic Review},
  volume={110},
  number={9},
  pages={2964--96},
  year={2020}
}

@techreport{goodman2018difference,
  title={Difference-in-differences with variation in treatment timing},
  author={Goodman-Bacon, Andrew},
  year={2018},
  institution={National Bureau of Economic Research}
}

@article{sun2020estimating,
  title={Estimating dynamic treatment effects in event studies with heterogeneous treatment effects},
  author={Sun, Liyang and Abraham, Sarah},
  journal={Journal of Econometrics},
  year={2020},
  publisher={Elsevier}
}

@article{wooldridge2005violating,
  title={Violating ignorability of treatment by controlling for too many factors},
  author={Wooldridge, Jeffrey M},
  journal={Econometric Theory},
  pages={1026--1028},
  year={2005},
  publisher={JSTOR}
}

@article{zimmert2020efficient,
      title={Efficient Difference-in-Differences Estimation with High-Dimensional Common Trend Confounding}, 
      author={Michael Zimmert},
      year={2020},
      eprint={1809.01643},
      archivePrefix={arXiv},
      journal={arXiv:1809.01643 },
      primaryClass={econ.EM}
}

@article{zimmert2018difference,
  title={Difference-in-differences estimation with high-dimensional common trend confounding},
  author={Zimmert, Michael},
  journal={arXiv preprint arXiv:1809.01643},
  year={2018}
}

@article{hong2013measuring,
  title={Measuring the effect of napster on recorded music sales: difference-in-differences estimates under compositional changes},
  author={Hong, Seung-Hyun},
  journal={Journal of Applied Econometrics},
  volume={28},
  number={2},
  pages={297--324},
  year={2013},
  publisher={Wiley Online Library}
}

@article{abadie2005semiparametric,
  title={Semiparametric difference-in-differences estimators},
  author={Abadie, Alberto},
  journal={The review of economic studies},
  volume={72},
  number={1},
  pages={1--19},
  year={2005},
  publisher={Wiley-Blackwell}
}

@article{chang2020double,
  title={Double/debiased machine learning for difference-in-differences models},
  author={Chang, Neng-Chieh},
  journal={The Econometrics Journal},
  volume={23},
  number={2},
  pages={177--191},
  year={2020},
  publisher={Oxford University Press}
}

@article{lu2019robust,
  title={Robust nonparametric difference-in-differences estimation},
  author={Lu, Chen and Nie, Xinkun and Wager, Stefan},
  journal={arXiv preprint arXiv:1905.11622},
  year={2019}
}

@article{macinko2015brazil,
  title={Brazil’s family health strategy—delivering community-based primary care in a universal health system},
  author={Macinko, James and Harris, Matthew J},
  journal={N Engl J Med},
  volume={372},
  number={23},
  pages={2177--81},
  year={2015}
}

@article{hone2017association,
  title={Association between expansion of primary healthcare and racial inequalities in mortality amenable to primary care in Brazil: A national longitudinal analysis},
  author={Hone, Thomas and Rasella, Davide and Barreto, Mauricio L and Majeed, Azeem and Millett, Christopher},
  journal={PLoS medicine},
  volume={14},
  number={5},
  pages={e1002306},
  year={2017},
  publisher={Public Library of Science San Francisco, CA USA}
}

@article{aquino2009impact,
  title={Impact of the family health program on infant mortality in Brazilian municipalities},
  author={Aquino, Rosana and De Oliveira, Nelson F and Barreto, Mauricio L},
  journal={American journal of public health},
  volume={99},
  number={1},
  pages={87--93},
  year={2009},
  publisher={American Public Health Association}
}

@article{castro2019brazil,
  title={Brazil's unified health system: the first 30 years and prospects for the future},
  author={Castro, Marcia C and Massuda, Adriano and Almeida, Gisele and Menezes-Filho, Naercio Aquino and Andrade, Monica Viegas and de Souza Noronha, Kenya Val{\'e}ria Micaela and Rocha, Rudi and Macinko, James and Hone, Thomas and Tasca, Renato and others},
  journal={The Lancet},
  volume={394},
  number={10195},
  pages={345--356},
  year={2019},
  publisher={Elsevier}
}

@article{dourado2011trends,
  title={Trends in primary health care-sensitive conditions in Brazil: the role of the Family Health Program (Project ICSAP-Brazil)},
  author={Dourado, Ines and Oliveira, Veneza B and Aquino, Rosana and Bonolo, Palmira and Lima-Costa, Maria Fernanda and Medina, Maria Guadalupe and Mota, Eduardo and Turci, Maria A and Macinko, James},
  journal={Medical care},
  pages={577--584},
  year={2011},
  publisher={JSTOR}
}

@article{de2016social,
  title={Social inequalities in health behaviors among Brazilian adults: National Health Survey, 2013},
  author={de Azevedo Barros, Marilisa Berti and Lima, Margareth Guimar{\~a}es and Medina, Lhais de Paula Barbosa and Szwarcwald, Celia Landman and Malta, Deborah Carvalho},
  journal={International journal for equity in health},
  volume={15},
  number={1},
  pages={1--10},
  year={2016},
  publisher={Springer}
}

@article{rocha2010evaluating,
  title={Evaluating the impact of community-based health interventions: evidence from Brazil's Family Health Program},
  author={Rocha, Romero and Soares, Rodrigo R},
  journal={Health Economics},
  volume={19},
  number={S1},
  pages={126--158},
  year={2010},
  publisher={Wiley Online Library}
}

@article{bhalotra2016does,
  title={Does universalization of healthwork? Evidence from health systems restructuring and expansion in Brazil},
  author={Bhalotra, Sonia R and Rocha, Rudi and Soares, Rodrigo R},
  year={2019},
  journal={IZA Discussion Paper},
volume={12111}
}
%\bibliographystyle{agsm}
%\bibliography{GEJM}

\appendix

\setcounter{table}{0}
\renewcommand{\thetable}{A\arabic{table}}

\setcounter{figure}{0}
\renewcommand{\thefigure}{A\arabic{figure}}

\newpage

\begin{table}
\begin{center}
\caption{Simulation Results: Average Group-Time ATTs (no confounding)}
\begin{adjustbox}{width=0.8\textwidth}
\begin{tabular}{l c c c | c c c | c c c}
\hline
& \multicolumn{9}{c}{H0C0B0X0, tau by X1} \\ 
& \multicolumn{3}{c}{N=2500} & \multicolumn{3}{c}{N=5000} & \multicolumn{3}{c}{N=10000} \\
 & Oracle & MLDID & DRDID & Oracle & MLDID & DRDID & Oracle & MLDID & DRDID \\
\hline
t1   & $0.00$   & $0.00$   & $0.00$   & $0.00$   & $0.00$   & $0.00$   & $0.00$   & $0.00$   & $0.00$   \\
t2   & $-3.99$  & $-4.01$  & $-4.00$  & $-4.00$  & $-4.02$  & $-4.01$  & $-4.00$  & $-4.02$  & $-4.00$  \\
     & $(0.00)$ & $(0.12)$ & $(0.14)$ & $(0.00)$ & $(0.09)$ & $(0.10)$ & $(0.00)$ & $(0.06)$ & $(0.07)$ \\
t3   & $-5.00$  & $-4.98$  & $-5.00$  & $-5.00$  & $-4.99$  & $-5.00$  & $-5.00$  & $-5.02$  & $-5.00$  \\
     & $(0.00)$ & $(0.14)$ & $(0.17)$ & $(0.00)$ & $(0.10)$ & $(0.12)$ & $(0.00)$ & $(0.07)$ & $(0.09)$ \\
t4   & $-5.99$  & $-5.98$  & $-5.99$  & $-6.00$  & $-6.00$  & $-6.01$  & $-6.00$  & $-6.00$  & $-6.00$  \\
     & $(0.00)$ & $(0.14)$ & $(0.22)$ & $(0.00)$ & $(0.10)$ & $(0.15)$ & $(0.00)$ & $(0.07)$ & $(0.11)$ \\
     \hdashline
t1.1 & $0.00$   & $-0.00$  & $-0.01$  & $0.00$   & $0.01$   & $0.00$   & $0.00$   & $-0.00$  & $-0.00$  \\
     & $(0.00)$ & $(0.15)$ & $(0.08)$ & $(0.00)$ & $(0.10)$ & $(0.05)$ & $(0.00)$ & $(0.07)$ & $(0.04)$ \\
t1.2 & $0.00$   & $0.00$   & $0.00$   & $0.00$   & $0.00$   & $0.00$   & $0.00$   & $0.00$   & $0.00$   \\
t1.3 & $-4.99$  & $-5.18$  & $-4.99$  & $-5.01$  & $-5.03$  & $-5.01$  & $-5.01$  & $-5.22$  & $-5.01$  \\
     & $(0.00)$ & $(0.19)$ & $(0.14)$ & $(0.00)$ & $(0.12)$ & $(0.10)$ & $(0.00)$ & $(0.08)$ & $(0.07)$ \\
t1.4 & $-5.98$  & $-5.99$  & $-5.98$  & $-6.01$  & $-6.01$  & $-6.02$  & $-6.01$  & $-6.01$  & $-6.01$  \\
     & $(0.00)$ & $(0.14)$ & $(0.18)$ & $(0.00)$ & $(0.10)$ & $(0.13)$ & $(0.00)$ & $(0.07)$ & $(0.09)$ \\
     \hdashline
t2.1 & $0.00$   & $0.01$   & $0.01$   & $0.00$   & $0.01$   & $0.01$   & $0.00$   & $-0.00$  & $0.00$   \\
     & $(0.00)$ & $(0.15)$ & $(0.09)$ & $(0.00)$ & $(0.10)$ & $(0.06)$ & $(0.00)$ & $(0.07)$ & $(0.04)$ \\
t2.2 & $0.00$   & $0.02$   & $0.01$   & $0.00$   & $0.01$   & $0.01$   & $0.00$   & $-0.02$  & $-0.00$  \\
     & $(0.00)$ & $(0.18)$ & $(0.09)$ & $(0.00)$ & $(0.13)$ & $(0.06)$ & $(0.00)$ & $(0.09)$ & $(0.05)$ \\
t2.3 & $0.00$   & $0.00$   & $0.00$   & $0.00$   & $0.00$   & $0.00$   & $0.00$   & $0.00$   & $0.00$   \\
t2.4 & $-5.98$  & $-5.96$  & $-5.97$  & $-5.99$  & $-5.99$  & $-5.99$  & $-6.01$  & $-6.02$  & $-6.01$  \\
     & $(0.00)$ & $(0.13)$ & $(0.15)$ & $(0.00)$ & $(0.09)$ & $(0.10)$ & $(0.00)$ & $(0.06)$ & $(0.07)$ \\
\hline
& \multicolumn{9}{c}{H0C0B1X1, tau by X1} \\ 
& \multicolumn{3}{c}{N=2500} & \multicolumn{3}{c}{N=5000} & \multicolumn{3}{c}{N=10000} \\
 & Oracle & MLDID & DRDID & Oracle & MLDID & DRDID & Oracle & MLDID & DRDID \\
\hline
t1   & $0.00$   & $0.00$   & $0.00$   & $0.00$   & $0.00$   & $0.00$   & $0.00$   & $0.00$   & $0.00$   \\
     & $(0.00)$ & $(0.00)$ & $(0.00)$ & $(0.00)$ & $(0.00)$ & $(0.00)$ & $(0.00)$ & $(0.00)$ & $(0.00)$ \\
t2   & $-3.99$  & $-4.01$  & $-4.00$  & $-4.00$  & $-4.02$  & $-4.01$  & $-4.00$  & $-4.02$  & $-4.00$  \\
     & $(0.00)$ & $(0.12)$ & $(0.14)$ & $(0.00)$ & $(0.09)$ & $(0.10)$ & $(0.00)$ & $(0.06)$ & $(0.07)$ \\
t3   & $-5.00$  & $-4.98$  & $-5.00$  & $-5.00$  & $-4.99$  & $-5.00$  & $-5.00$  & $-5.02$  & $-5.00$  \\
     & $(0.00)$ & $(0.14)$ & $(0.17)$ & $(0.00)$ & $(0.10)$ & $(0.12)$ & $(0.00)$ & $(0.07)$ & $(0.09)$ \\
t4   & $-5.99$  & $-5.98$  & $-5.99$  & $-6.00$  & $-6.00$  & $-6.01$  & $-6.00$  & $-6.00$  & $-6.00$  \\
     & $(0.00)$ & $(0.14)$ & $(0.22)$ & $(0.00)$ & $(0.10)$ & $(0.15)$ & $(0.00)$ & $(0.07)$ & $(0.11)$ \\
t1.1 & $0.00$   & $-0.00$  & $-0.01$  & $0.00$   & $0.01$   & $0.00$   & $0.00$   & $-0.00$  & $-0.00$  \\
     & $(0.00)$ & $(0.15)$ & $(0.08)$ & $(0.00)$ & $(0.10)$ & $(0.05)$ & $(0.00)$ & $(0.07)$ & $(0.04)$ \\
t1.2 & $0.00$   & $0.00$   & $0.00$   & $0.00$   & $0.00$   & $0.00$   & $0.00$   & $0.00$   & $0.00$   \\
     & $(0.00)$ & $(0.00)$ & $(0.00)$ & $(0.00)$ & $(0.00)$ & $(0.00)$ & $(0.00)$ & $(0.00)$ & $(0.00)$ \\
t1.3 & $-4.99$  & $-5.18$  & $-4.99$  & $-5.01$  & $-5.03$  & $-5.01$  & $-5.01$  & $-5.22$  & $-5.01$  \\
     & $(0.00)$ & $(0.19)$ & $(0.14)$ & $(0.00)$ & $(0.12)$ & $(0.10)$ & $(0.00)$ & $(0.08)$ & $(0.07)$ \\
t1.4 & $-5.98$  & $-5.99$  & $-5.98$  & $-6.01$  & $-6.01$  & $-6.02$  & $-6.01$  & $-6.01$  & $-6.01$  \\
     & $(0.00)$ & $(0.14)$ & $(0.18)$ & $(0.00)$ & $(0.10)$ & $(0.13)$ & $(0.00)$ & $(0.07)$ & $(0.09)$ \\
t2.1 & $0.00$   & $0.01$   & $0.01$   & $0.00$   & $0.01$   & $0.01$   & $0.00$   & $-0.00$  & $0.00$   \\
     & $(0.00)$ & $(0.15)$ & $(0.09)$ & $(0.00)$ & $(0.10)$ & $(0.06)$ & $(0.00)$ & $(0.07)$ & $(0.04)$ \\
t2.2 & $0.00$   & $0.02$   & $0.01$   & $0.00$   & $0.01$   & $0.01$   & $0.00$   & $-0.02$  & $-0.00$  \\
     & $(0.00)$ & $(0.18)$ & $(0.09)$ & $(0.00)$ & $(0.13)$ & $(0.06)$ & $(0.00)$ & $(0.09)$ & $(0.05)$ \\
t2.3 & $0.00$   & $0.00$   & $0.00$   & $0.00$   & $0.00$   & $0.00$   & $0.00$   & $0.00$   & $0.00$   \\
     & $(0.00)$ & $(0.00)$ & $(0.00)$ & $(0.00)$ & $(0.00)$ & $(0.00)$ & $(0.00)$ & $(0.00)$ & $(0.00)$ \\
t2.4 & $-5.98$  & $-5.96$  & $-5.97$  & $-5.99$  & $-5.99$  & $-5.99$  & $-6.01$  & $-6.02$  & $-6.01$  \\
     & $(0.00)$ & $(0.13)$ & $(0.15)$ & $(0.00)$ & $(0.09)$ & $(0.10)$ & $(0.00)$ & $(0.06)$ & $(0.07)$ \\
\hline
\end{tabular}
%\caption{Statistical models}
\label{table:coefficients}
\end{adjustbox}
\end{center}
\end{table}

\begin{table}
\begin{center}
\caption{Simulation Results: Average Group-Time ATTs (confounding in X2)}
\begin{adjustbox}{width=0.8\textwidth}
\begin{tabular}{l c c c | c c c | c c c}
\hline
& \multicolumn{9}{c}{$\tau$ by X1} \\ 
& \multicolumn{3}{c}{N=2500} & \multicolumn{3}{c}{N=5000} & \multicolumn{3}{c}{N=10000} \\
 & Oracle & MLDID & DRDID & Oracle & MLDID & DRDID & Oracle & MLDID & DRDID \\
\hline
t1   & $0.00$   & $0.00$   & $0.00$   & $0.00$   & $0.00$   & $0.00$   & $0.00$   & $0.00$   & $0.00$   \\
t2   & $-3.99$  & $-3.99$  & $-3.99$  & $-4.00$  & $-3.99$  & $-3.99$  & $-4.01$  & $-4.01$  & $-4.01$  \\
     & $(0.00)$ & $(0.12)$ & $(0.14)$ & $(0.00)$ & $(0.09)$ & $(0.10)$ & $(0.00)$ & $(0.06)$ & $(0.07)$ \\
t3   & $-4.99$  & $-5.00$  & $-4.99$  & $-4.99$  & $-5.01$  & $-5.00$  & $-5.02$  & $-5.01$  & $-5.02$  \\
     & $(0.00)$ & $(0.14)$ & $(0.18)$ & $(0.00)$ & $(0.10)$ & $(0.12)$ & $(0.00)$ & $(0.07)$ & $(0.09)$ \\
t4   & $-5.98$  & $-5.98$  & $-5.98$  & $-5.99$  & $-5.99$  & $-6.00$  & $-6.02$  & $-6.02$  & $-6.02$  \\
     & $(0.00)$ & $(0.14)$ & $(0.22)$ & $(0.00)$ & $(0.10)$ & $(0.15)$ & $(0.00)$ & $(0.07)$ & $(0.11)$ \\
     \hdashline
t1.1 & $0.00$   & $0.01$   & $-0.00$  & $0.00$   & $0.00$   & $0.01$   & $0.00$   & $-0.01$  & $-0.01$  \\
     & $(0.00)$ & $(0.15)$ & $(0.08)$ & $(0.00)$ & $(0.10)$ & $(0.06)$ & $(0.00)$ & $(0.07)$ & $(0.04)$ \\
t1.2 & $0.00$   & $0.00$   & $0.00$   & $0.00$   & $0.00$   & $0.00$   & $0.00$   & $0.00$   & $0.00$   \\
t1.3 & $-5.00$  & $-5.18$  & $-5.00$  & $-5.00$  & $-5.19$  & $-5.00$  & $-5.01$  & $-5.22$  & $-5.01$  \\
     & $(0.00)$ & $(0.18)$ & $(0.14)$ & $(0.00)$ & $(0.11)$ & $(0.10)$ & $(0.00)$ & $(0.08)$ & $(0.07)$ \\
t1.4 & $-5.99$  & $-5.98$  & $-5.99$  & $-6.02$  & $-5.98$  & $-6.01$  & $-6.01$  & $-6.00$  & $-6.01$  \\
     & $(0.00)$ & $(0.14)$ & $(0.18)$ & $(0.00)$ & $(0.10)$ & $(0.13)$ & $(0.00)$ & $(0.07)$ & $(0.09)$ \\
     \hdashline
t2.1 & $0.00$   & $0.01$   & $0.01$   & $0.00$   & $-0.01$  & $-0.00$  & $0.00$   & $0.01$   & $0.01$   \\
     & $(0.00)$ & $(0.15)$ & $(0.09)$ & $(0.00)$ & $(0.11)$ & $(0.07)$ & $(0.00)$ & $(0.07)$ & $(0.05)$ \\
t2.2 & $0.00$   & $-0.02$  & $0.02$   & $0.00$   & $-0.02$  & $-0.00$  & $0.00$   & $-0.03$  & $-0.01$  \\
     & $(0.00)$ & $(0.19)$ & $(0.09)$ & $(0.00)$ & $(0.13)$ & $(0.07)$ & $(0.00)$ & $(0.09)$ & $(0.05)$ \\
t2.3 & $0.00$   & $0.00$   & $0.00$   & $0.00$   & $0.00$   & $0.00$   & $0.00$   & $0.00$   & $0.00$   \\
t2.4 & $-6.01$  & $-5.96$  & $-6.00$  & $-6.00$  & $-5.97$  & $-6.00$  & $-6.00$  & $-5.98$  & $-6.00$  \\
     & $(0.00)$ & $(0.13)$ & $(0.15)$ & $(0.00)$ & $(0.09)$ & $(0.11)$ & $(0.00)$ & $(0.07)$ & $(0.08)$ \\
\hline
\hline
& \multicolumn{9}{c}{$\tau$ by $(X2+X3)^2$} \\ 
& \multicolumn{3}{c}{N=2500} & \multicolumn{3}{c}{N=5000} & \multicolumn{3}{c}{N=10000} \\
 & Oracle & MLDID & DRDID & Oracle & MLDID & DRDID & Oracle & MLDID & DRDID \\
\hline
t1   & $0.00$   & $0.00$   & $0.00$   & $0.00$   & $0.00$   & $0.00$   & $0.00$   & $0.00$   & $0.00$   \\
t2   & $-1.07$  & $-0.97$  & $-1.07$  & $-1.05$  & $-0.96$  & $-1.04$  & $-1.06$  & $-0.97$  & $-1.06$  \\
     & $(0.00)$ & $(0.21)$ & $(0.21)$ & $(0.00)$ & $(0.15)$ & $(0.15)$ & $(0.00)$ & $(0.10)$ & $(0.11)$ \\
t3   & $-0.61$  & $-0.59$  & $-0.61$  & $-0.57$  & $-0.57$  & $-0.58$  & $-0.59$  & $-0.56$  & $-0.59$  \\
     & $(0.00)$ & $(0.28)$ & $(0.29)$ & $(0.00)$ & $(0.20)$ & $(0.21)$ & $(0.00)$ & $(0.14)$ & $(0.15)$ \\
t4   & $-0.14$  & $-0.63$  & $-0.14$  & $-0.10$  & $-0.57$  & $-0.10$  & $-0.11$  & $-0.60$  & $-0.11$  \\
     & $(0.00)$ & $(0.34)$ & $(0.38)$ & $(0.00)$ & $(0.24)$ & $(0.27)$ & $(0.00)$ & $(0.17)$ & $(0.19)$ \\
     \hdashline
t1.1 & $0.00$   & $0.01$   & $-0.00$  & $0.00$   & $0.00$   & $0.01$   & $0.00$   & $-0.01$  & $-0.01$  \\
     & $(0.00)$ & $(0.15)$ & $(0.08)$ & $(0.00)$ & $(0.10)$ & $(0.06)$ & $(0.00)$ & $(0.07)$ & $(0.04)$ \\
t1.2 & $0.00$   & $0.00$   & $0.00$   & $0.00$   & $0.00$   & $0.00$   & $0.00$   & $0.00$   & $0.00$   \\
t1.3 & $-1.78$  & $-1.99$  & $-1.78$  & $-1.77$  & $-1.97$  & $-1.77$  & $-1.78$  & $-2.02$  & $-1.78$  \\
     & $(0.00)$ & $(0.26)$ & $(0.22)$ & $(0.00)$ & $(0.19)$ & $(0.16)$ & $(0.00)$ & $(0.13)$ & $(0.11)$ \\
t1.4 & $-1.15$  & $-1.87$  & $-1.16$  & $-1.17$  & $-1.87$  & $-1.16$  & $-1.16$  & $-1.91$  & $-1.16$  \\
     & $(0.00)$ & $(0.30)$ & $(0.32)$ & $(0.00)$ & $(0.22)$ & $(0.23)$ & $(0.00)$ & $(0.15)$ & $(0.16)$ \\
     \hdashline
t2.1 & $0.00$   & $0.01$   & $0.01$   & $0.00$   & $-0.01$  & $-0.00$  & $0.00$   & $0.01$   & $0.01$   \\
     & $(0.00)$ & $(0.15)$ & $(0.09)$ & $(0.00)$ & $(0.11)$ & $(0.07)$ & $(0.00)$ & $(0.07)$ & $(0.05)$ \\
t2.2 & $0.00$   & $-0.02$  & $0.02$   & $0.00$   & $-0.02$  & $-0.00$  & $0.00$   & $-0.03$  & $-0.01$  \\
     & $(0.00)$ & $(0.19)$ & $(0.09)$ & $(0.00)$ & $(0.13)$ & $(0.07)$ & $(0.00)$ & $(0.09)$ & $(0.05)$ \\
t2.3 & $0.00$   & $0.00$   & $0.00$   & $0.00$   & $0.00$   & $0.00$   & $0.00$   & $0.00$   & $0.00$   \\
t2.4 & $-2.46$  & $-3.21$  & $-2.46$  & $-2.47$  & $-3.20$  & $-2.47$  & $-2.46$  & $-3.19$  & $-2.46$  \\
     & $(0.00)$ & $(0.22)$ & $(0.24)$ & $(0.00)$ & $(0.16)$ & $(0.17)$ & $(0.00)$ & $(0.11)$ & $(0.12)$ \\
     \hline
     \hline
\end{tabular}
%\caption{Statistical models}
\label{table:coefficients}
\end{adjustbox}
\end{center}
\end{table}

%\def\sym#1{\ifmmode^{#1}\else\(^{#1}\)\fi}
%\begin{landscape}
\begin{table}[htbp]
%\begin{table}[htbp] 
    \centering
    \caption{Summary Statistics (Outcomes): Main Sample (1995-2012)\tnote{a}} 
    \begin{adjustbox}{width=\textwidth}
    \begin{threeparttable}
    \begin{tabular}{lccccc} 
    \toprule
            \multicolumn{5}{c}{Main Outcomes of Interest\tnote{a}}                                            \\
            &        mean&          sd&         min&         max&       count\\
    \midrule 
Rate of births per woman (age 10-49)&        0.05&        0.03&        0.00&        0.18&        4972\\
Total mortality rate (per 1000 population)&        4.95&        2.34&        0.03&       13.25&        4931\\
Female mortality rate (per 1000 age 10-49)&        1.01&        0.80&        0.00&        6.86&        4972\\
Maternal mortality (per 1000 live births)&        2.52&       32.06&        0.00&     1000.00&        3826\\
Infant mortality (per 1000 age 0-1)&       18.62&       17.95&        0.00&      221.61&        4969\\
Infant mortality, within 24 hours&        3.88&        6.02&        0.00&       90.91&        4894\\
Infant mortality, neonatal&        6.21&        7.67&        0.00&       75.47&        4946\\
Infant mortality, 1 month to 1 year&        8.69&       11.23&        0.00&      163.43&        4924\\
Fetal mortality (per 1000 age 0-1)&        8.67&       10.77&        0.00&      142.86&        4962\\
    \midrule
            %&\multicolumn{5}{c}{(1)}                                         \\
    \midrule 
            &        mean&          sd&         min&         max&       count\\
\hline
\textbf{Adult Mortality, by cause (per 1000 population)} & & & & & \\
Circulatory         &        1.31&        1.03&        0.00&        9.28&        4931\\
Respiratory         &        0.44&        0.41&        0.00&        3.48&        4931\\
Neoplasms           &        0.49&        0.45&        0.00&        2.90&        4931\\
Diabetes            &        0.11&        0.14&        0.00&        1.51&        4972\\
Hypertension        &        0.10&        0.15&        0.00&        1.99&        4972\\
\textbf{Infant Mortality, by cause (per 1000 age 0-1)} & & & & & \\
Infectious          &        2.50&        4.35&        0.00&       42.90&        4262\\
Respiratory         &        1.73&        3.21&        0.00&       32.26&        4236\\
Perinatal           &        8.07&        9.37&        0.00&       90.91&        4959\\
Congenital          &        1.68&        3.83&        0.00&       54.05&        4827\\
    \bottomrule
    \end{tabular}
    \begin{tablenotes}
        \item[a] Birth rate taken from Datasus/SINASC. Mortality information taken from Datasus/SIM and ICD-10 categorization of causes of death. For 1995, causes were classified by ICD-9 and therefore manually converted to match categorizations after 1996 (ICD-10). 
    \end{tablenotes}
    \end{threeparttable}
    \end{adjustbox}
\end{table}
%\end{landscape}
%\end{table}

\begin{table}
\caption{Effect of PSF on Infant Mortality: Dynamic ATTs}
\begin{center}
\begin{tabular}{l c c}
\hline
 Time & MLDiD & DRDiD \\
\hline
0 & $-2.90$  & $-4.07$  \\
  & $(2.53)$ & $(1.38)$ \\
1 & $-4.79$  & $-7.46$  \\
  & $(2.39)$ & $(1.76)$ \\
2 & $-5.57$  & $-6.74$  \\
  & $(2.34)$ & $(2.12)$ \\
3 & $-7.37$  & $-7.34$  \\
  & $(2.32)$ & $(2.48)$ \\
4 & $-10.05$ & $-7.77$  \\
  & $(2.52)$ & $(2.67)$ \\
5 & $-10.13$ & $-8.48$  \\
  & $(2.72)$ & $(2.92)$ \\
6 & $-15.40$ & $-9.34$  \\
  & $(3.42)$ & $(3.21)$ \\
7 & $-13.38$ & $-10.82$ \\
  & $(4.02)$ & $(3.59)$ \\
8 & $-19.50$ & $-14.51$ \\
  & $(5.24)$ & $(4.37)$ \\
\hline
\end{tabular}
%\caption{Effect of PSF on Infant Mortality}
\label{table:dynamiccoefficients}
\end{center}
\end{table}

\begin{table}[!htbp] \centering 
  \caption{Dynamic Best Linear Predictors of Heterogeneity} 
  \label{table:BLPdynamicPSF} 
  \begin{adjustbox}{max width=0.9\textwidth, max height=0.8\textheight}
\begin{tabular}{@{\extracolsep{5pt}}lccccccccc} 
\\[-1.8ex]\hline 
\hline \\[-1.8ex] 
 & \multicolumn{9}{c}{\textit{Dependent variable: score}} \\ 
\cline{2-10} 
\\[-1.8ex] & \multicolumn{8}{c}{Time Since Event} \\ 
\\[-1.8ex] & (0) & (1) & (2) & (3) & (4) & (5) & (6) & (7) & (8)\\ 
\hline \\[-1.8ex] 
Gini Index & $-$4.541 & $-$11.137 & 6.941 & $-$3.956 & 5.159 & 4.426 & 12.451 & 18.963 & 13.727 \\ 
  & (13.626) & (12.786) & (13.577) & (13.723) & (13.664) & (13.336) & (17.706) & (18.369) & (25.899) \\ 
  & & & & & & & & & \\ 
 \% Urban & $-$0.753 & $-$1.329 & $-$1.106 & $-$0.605 & 0.605 & $-$2.055$^{***}$ & $-$5.575$^{***}$ & $-$3.928$^{***}$ & $-$4.922$^{***}$ \\ 
  & (1.008) & (0.936) & (0.973) & (0.808) & (0.787) & (0.764) & (0.996) & (1.050) & (1.501) \\ 
  & & & & & & & & & \\ 
 Income PC & 0.00000 & 0.00000 & $-$0.00000 & 0.006 & 0.012$^{**}$ & 0.001 & 0.007 & $-$0.019$^{**}$ & $-$0.027$^{***}$ \\ 
  & (0.00000) & (0.00000) & (0.00000) & (0.007) & (0.006) & (0.006) & (0.007) & (0.008) & (0.010) \\ 
  & & & & & & & & & \\ 
 Poverty (140/day) & $-$25.932$^{**}$ & $-$27.353$^{**}$ & $-$36.997$^{***}$ & $-$24.441$^{**}$ & $-$10.735 & $-$39.431$^{***}$ & $-$48.770$^{***}$ & $-$61.817$^{***}$ & $-$68.525$^{***}$ \\ 
  & (10.849) & (11.165) & (12.237) & (12.375) & (12.347) & (12.209) & (16.027) & (16.362) & (22.763) \\ 
  & & & & & & & & & \\ 
 \% Literate & $-$19.127 & $-$31.593$^{**}$ & $-$40.122$^{**}$ & $-$27.018$^{*}$ & $-$57.278$^{***}$ & $-$43.538$^{***}$ & $-$33.212$^{*}$ & $-$54.755$^{***}$ & $-$83.483$^{***}$ \\ 
  & (15.872) & (15.885) & (17.495) & (15.094) & (15.060) & (14.780) & (19.490) & (20.185) & (28.808) \\ 
  & & & & & & & & & \\ 
 \% Black & $-$2.498 & $-$14.851 & $-$21.071 & $-$17.929 & $-$41.370$^{***}$ & $-$35.085$^{***}$ & $-$45.731$^{***}$ & $-$49.594$^{***}$ & $-$77.161$^{***}$ \\ 
  & (16.521) & (15.232) & (15.670) & (12.892) & (12.516) & (11.795) & (15.443) & (15.652) & (22.387) \\ 
  & & & & & & & & & \\ 
 \% Pardo & 2.762 & $-$2.823 & $-$4.714 & $-$3.603 & $-$13.453$^{**}$ & $-$4.880 & 0.754 & $-$10.525 & $-$23.254$^{**}$ \\ 
  & (6.131) & (5.825) & (6.180) & (5.286) & (5.232) & (5.085) & (6.661) & (6.783) & (9.613) \\ 
  & & & & & & & & & \\ 
 \% Indigenous & $-$0.109 & 4.065 & $-$3.545 & 2.430 & $-$15.052 & 12.402 & 11.560 & 18.801 & 4.536 \\ 
  & (24.459) & (22.032) & (22.164) & (17.560) & (16.609) & (15.069) & (18.608) & (19.611) & (25.320) \\ 
  & & & & & & & & & \\ 
 Hospital & $-$2.095 & $-$1.957 & $-$2.844$^{*}$ & $-$0.755 & $-$1.192 & $-$1.967 & $-$1.182 & $-$0.871 & 0.726 \\ 
  & (1.750) & (1.584) & (1.621) & (1.323) & (1.271) & (1.198) & (1.529) & (1.549) & (2.206) \\ 
  & & & & & & & & & \\ 
 \% Secondary Education & 17.402 & 17.258 & $-$1.323 & 1.162 & 6.847 & 46.621$^{***}$ & 81.458$^{***}$ & 92.081$^{***}$ & 108.885$^{***}$ \\ 
  & (17.091) & (16.185) & (16.839) & (15.020) & (14.072) & (12.525) & (16.086) & (17.420) & (24.585) \\ 
  & & & & & & & & & \\ 
 Proportion ESF & $-$0.007 & $-$0.042$^{**}$ & $-$0.023 & 0.018 & 0.033$^{*}$ & 0.021 & 0.005 & 0.005 & $-$0.022 \\ 
  & (0.025) & (0.021) & (0.022) & (0.019) & (0.019) & (0.018) & (0.024) & (0.026) & (0.037) \\ 
  & & & & & & & & & \\ 
 NE & 0.045 & 0.973 & 0.481 & $-$0.854 & $-$1.246 & $-$1.716 & $-$0.509 & 1.679 & $-$0.974 \\ 
  & (3.095) & (2.781) & (2.850) & (2.311) & (2.270) & (2.392) & (3.100) & (3.056) & (4.137) \\ 
  & & & & & & & & & \\ 
 CO & $-$4.356 & $-$3.287 & $-$4.709 & $-$4.780$^{*}$ & $-$2.571 & $-$8.712$^{***}$ & $-$10.134$^{***}$ & $-$5.070 & $-$5.695 \\ 
  & (3.701) & (3.414) & (3.507) & (2.790) & (2.675) & (2.635) & (3.333) & (3.281) & (4.571) \\ 
  & & & & & & & & & \\ 
 SoE & $-$10.635$^{***}$ & $-$16.140$^{***}$ & $-$14.025$^{***}$ & $-$12.774$^{***}$ & $-$10.919$^{***}$ & $-$17.404$^{***}$ & $-$20.802$^{***}$ & $-$15.974$^{***}$ & $-$20.639$^{***}$ \\ 
  & (3.566) & (3.262) & (3.323) & (2.608) & (2.512) & (2.501) & (3.173) & (3.065) & (4.067) \\ 
  & & & & & & & & & \\ 
 S & $-$6.602 & $-$6.058 & $-$7.011$^{*}$ & $-$8.252$^{***}$ & $-$9.451$^{***}$ & $-$14.809$^{***}$ & $-$18.765$^{***}$ & $-$12.546$^{***}$ & $-$17.280$^{***}$ \\ 
  & (4.121) & (3.786) & (3.887) & (3.114) & (2.998) & (2.923) & (3.734) & (3.659) & (4.976) \\ 
  & & & & & & & & & \\ 
 Constant & 24.684 & 43.870$^{***}$ & 48.779$^{***}$ & 32.693$^{**}$ & 48.556$^{***}$ & 43.225$^{***}$ & 21.045 & 47.337$^{**}$ & 85.176$^{***}$ \\ 
  & (15.339) & (15.262) & (16.732) & (14.390) & (14.372) & (14.388) & (18.920) & (19.134) & (27.089) \\ 
  & & & & & & & & & \\ 
\hline \\[-1.8ex] 
Observations & 4,360 & 4,323 & 4,287 & 4,177 & 3,948 & 3,555 & 3,353 & 2,967 & 2,277 \\ 
R$^{2}$ & 0.006 & 0.019 & 0.015 & 0.014 & 0.025 & 0.039 & 0.052 & 0.044 & 0.047 \\ 
Adjusted R$^{2}$ & 0.002 & 0.016 & 0.011 & 0.011 & 0.021 & 0.035 & 0.048 & 0.039 & 0.040 \\ 
\hline 
\hline \\[-1.8ex] 
\textit{Note:} & \multicolumn{8}{r}{$^{*}$p$<$0.1; $^{**}$p$<$0.05; $^{***}$p$<$0.01} \\ 
\end{tabular}
\end{adjustbox}
\end{table}

\begin{table}[ht]
\caption{Dynamic CLANs, PSF Case Study}
\label{table:CLANdynamicPSF} 
    \begin{adjustbox}{width=0.9\textwidth}
\centering
\begin{tabular}{rllllllll}
  \hline
event-time & 1 & 2 & 3 & 4 & 5 & 6 & 7 & 8 \\ 
  \hline
Literacy &      -0.0114 &      -0.0475 &     -0.0356 &  -0.02497 &  -0.01646 &  0.0036 &  0.0242*** &  0.03331*** \\ 
&\\
  Black &       0.0033 &       0.0021 &      0.0041* &   0.00224 &  -0.00333 &  0.0037 & -0.0023 &  0.00098 \\ 
  &\\
  Poverty &      -0.0024 &       0.0713*** &      0.0695*** &   0.04629*** &   0.02191** & -0.0034 & -0.0319 & -0.04676 \\ 
  &\\
  SE Region &      -0.0367 &      -0.3249 &     -0.2727 &  -0.12081 &   0.07468** &  0.0872*** &  0.0611* &  0.12121*** \\ 
  &\\
  S &      -0.0069 &       0.0925*** &      0.0816*** &   0.02871* &  -0.06329 & -0.0872 &  0.0030 & -0.01010 \\ 
  &\\
  Gini &       0.0038 &       0.0157*** &      0.0166*** &   0.01350*** &   0.00989*** &  0.0092** & -0.0012 & -0.00375 \\ 
  &\\
  Central &       0.0057 &       0.0393*** &      0.0268** &   0.02153* &   0.02152* &  0.0197 &  0.0298* &  0.02862* \\ 
  &\\
  Pardo &       0.0269** &       0.0664*** &      0.0472*** &   0.01406 &  -0.00917 &  0.0022 & -0.0359 & -0.04017 \\ 
  &\\
  Indigenous &      -0.0011 &       0.0011 &      0.0017 &   0.00096 &  -0.00032 &  0.0023 &  0.0011 & -0.00136 \\ 
  &\\
  Hospital &       0.0115 &       0.0127 &      0.0152 &   0.01077 &   0.00633 &  0.0141 & -0.0358 & -0.06734 \\ 
  &\\
  Urban &      -0.0237 &      -0.4140 &     -0.3594 &  -0.18131 &   0.08297* &  0.0284 &  0.0220 &  0.09702* \\ 
  &\\
  NE Region&       0.0183 &       0.1561*** &      0.1177*** &   0.04187* &  -0.04557 & -0.0394 & -0.1058 & -0.15488 \\ 
  &\\
  Secondary Educ &      -0.0109 &      -0.0434 &     -0.0391 &  -0.03217 &  -0.02160 &  0.0041 &  0.0233*** &  0.02994*** \\ 
   \hline
\end{tabular}
\end{adjustbox}
\end{table}

\end{document}